%% file: main.tex
\documentclass{article}
\usepackage{fullpage}
\usepackage[T1]{fontenc}
\usepackage[utf8]{inputenc}
\usepackage{booktabs}

\usepackage{amsmath}
\usepackage{amsfonts}
\usepackage[figuresleft]{rotating}

\usepackage{bm}

\usepackage{enumerate}
\usepackage{graphicx}
\usepackage{subcaption}

\usepackage[super]{natbib} 

\usepackage[bookmarks]{hyperref} 
\hypersetup{
    colorlinks=true,
    linkcolor=black,
    citecolor=cyan,
    urlcolor=black,
}

\def \bsd {BSD}
\def \bsd {BSD}
\def \fooof {FOOOF}

\usepackage{authblk}

\title{\bsd: a Bayesian framework for parametric models of neural spectra
}

\author[1]{Johan Medrano}
\author[1]{Nicholas A. Alexander}
\author[1]{Robert A. Seymour}
\author[1]{Peter Zeidman}

\affil[1]{Functional Imaging Laboratory, Department of Imaging Neuroscience, UCL Queen Square Institute of Neurology, London, WC1N 3AR, UK}

\date{} 

\usepackage{lineno}
\usepackage{setspace}

\begin{document}
\maketitle

\begin{abstract}
The analysis of neural power spectra plays a crucial role in understanding brain function and dysfunction. While recent efforts have led to the development of methods for decomposing spectral data, challenges remain in performing statistical analysis and group-level comparisons. Here, we introduce Bayesian Spectral Decomposition (\bsd{}), a Bayesian framework for analysing neural spectral power. \bsd{} allows for the specification, inversion, comparison, and analysis of parametric models of neural spectra, addressing limitations of existing methods. We first establish the face validity of \bsd{} on simulated data and show how it outperforms an established method (\fooof{}) for peak detection on artificial spectral data. We then demonstrate the efficacy of \bsd{} on a group-level study of EEG spectra in 204 healthy subjects from the LEMON dataset. Our results not only highlight the effectiveness of \bsd{} in model selection and  parameter estimation, but also illustrate how \bsd{} enables straightforward group-level regression of the effect of continuous covariates such as age. By using Bayesian inference techniques, \bsd{} provides a robust framework for studying neural spectral data and their relationship to brain function and dysfunction.
\end{abstract}

\section{Introduction} 
    The spectral density of electrophysiological signals represents neurophysiological processes as a combination of rhythmic oscillations and aperiodic power with apparent scale-free behavior \cite{donoghue2020parameterizing,he2014scale}.  Rhythmic oscillations are assumed to play a fundamental role in neuronal computations \cite{buzsaki2012brain} and, as such, they are observed in electrophysiological measurements of most brain disorders and neuroimaging experiments \cite{buzsaki2004neuronal}. They have been hypothesised to enable communication between cortical populations \cite{fries2005mechanism}, thereby coordinating the emergence of centre manifolds that provide a stable scaffold on which cortical computations can take place \cite{huys2014functional}.  The remaining, aperiodic part of the spectrum, which exhibits a scale-free $1/f$ spectral density, also reflects neurophysiology \cite{he2014scale, allegrini2009spontaneous}. For instance, changes in the spectral aperiodic component related to healthy aging are well known \cite{hill2022periodic, thuwal2021aperiodic}, as well as in disorders such as ADHD \cite{karalunas2022electroencephalogram,arnett2022higher}. Both the periodic and aperiodic components are therefore relevant for clinical and cognitive neuroscience.

    Recent efforts have been focused on finding systematic approaches for decomposing spectral neural  data \cite{whitten2011better, wen2016separating}. One recent and popular method is Fitting Oscillations and One Over F (\fooof{}) \cite{donoghue2020parameterizing}. \fooof{} parameterises log power spectral densities as the sum of periodic and aperiodic components that are estimated from the data using an iterative procedure. Despite its relatively recent introduction, \fooof{} has already been successfully applied in hundreds of different studies, either to isolate and analyse the rhythmic oscillations \cite{trondle2023decomposing}, the aperiodic component \cite{merkin2023age}, to remove the confounding effect of the aperiodic component for group studies \cite{oswal2021neural}, or to improve the detection of transient oscillations \citep{seymour2022robust}. This highlights the strong relevance of the spectral model offered by \fooof{} for addressing research questions about brain function and disorder. 
    
    Despite being a successful tool for decomposing power spectra, \fooof{} may benefit from several improvements. For instance, \fooof{} necessitates selecting several parameters that can greatly influence the results. For this, there is no well-established approach to select parameters. In addition, there is no straightforward way to perform statistical analysis with \fooof{}; in particular, there is no standardised approach for performing group-level analysis. Some authors have used \fooof{} outputs to perform statistical analysis \citep{oswal2021neural}. However, these approaches fail to account for the uncertainty of parameters estimates: ideally, one would like to convey both estimated parameters and their confidence (precision) from the individual subjects to the group-level. 
    
    Fortunately, these limitations can be addressed by moving to a Bayesian framework. Bayesian statistics have already proven to be useful in other neuroimaging applications \cite{friston2002classical, friston2002classical2}. A range of tools for model-estimation, group-level analysis, and model selection have been devised and improved over the two last decades in neuroimaging \cite{ zeidman2023primer}. Specifically, we can leverage variational Bayesian inversion to estimate parameters \cite{friston2003dynamic, friston2007variational}, Bayesian Model Selection to choose the most adequate model \cite{kass1995bayes,stephan2009bayesian}, and Parametric Empirical Bayes (PEB) to conduct between-subject analysis while accounting for within-subject variability\cite{friston2016bayesian, zeidman2019guide2}. Overall, Bayesian methods enable standardised, integrated pipelines for performing statistical analysis of single-subject and group-subject data, which we can leverage for analysing neural spectra. 

    In this work, we introduce Bayesian Spectral Decomposition (\bsd{}): a Bayesian framework for specifying, estimating, comparing, and analysing parametric models of neural spectral power. This article is structured as follows. We first present the theoretical elements of \bsd{}, focusing on introducing only the immediately relevant concepts from Bayesian inference. We then proceed with establishing its face validity on simulated data, before showcasing its application in a large scale group-level study of the aging effect on the EEG spectra of 204 healthy subjects from the LEMON dataset. Finally, we discuss the potential applications \bsd{} enables and the future directions for its development. 
    
    \section{Theory}
        \subsection{Parametric spectral model}
            \bsd{} models the amplitude spectral density (here, "spectral density"), obtained by taking the square-root of power spectral density or the modulus of the Fourier coefficients of a signal. The power spectrum can computed using any approach, e.g.,  Fast-Fourier Transform, Welch's method, or using an autoregressive model. The spectrum is scaled to a variance of 8 to simplify model specification. The noise-free spectral density $S_f(\theta)$, for frequencies $f = (f_1,\dots,f_n)$ and some parameters $\theta$, is modelled as the sum of an aperiodic component ($A_f$) and $N$ periodic components ($P_{f,i}$): 
            \begin{align}
                S_f(\theta) = A_f(\theta) + \sum_{i=1}^N P_{f,i}(\theta) 
            \end{align}
            Each periodic component characterises a peak in the spectrum. Given that we are principally interested in the mean and width of a peak, a Gaussian model is suitable. Here, a single Gaussian peak has three parameters -- a mean $m_i$, a standard deviation $s_i$, and a height $h_i$, where $i$ indexes the peak. It has the following form: 
            \begin{align}
                P_{f,i}(\theta) = h_i(\theta) \exp\left( -\frac{1}{2} \left(\frac{f - m_i(\theta)}{s_i(\theta)}\right)^2\right)
                \label{eq:periodic}
            \end{align}
            The aperiodic component captures the $1/f^\alpha$ trend in power commonly observed in the spectrum of biophysical signals. It has the form: 
            \begin{align}
                A_f(\theta) = \gamma(\theta) f^{-\alpha(\theta)/2} 
                \label{eq:aperiodic}
            \end{align}
            where $\gamma$ is the broadband amplitude and $\alpha$ is the scaling exponent. The noise-free parametric spectral model, composed of the periodic and aperiodic components, forms the basis of the spectral likelihood. 
    
        \subsection{Spectral likelihood}    
            In practice, the observed spectrum $Y_f$ is corrupted by various sources of noise and unmodelled effects. We account for these by adding a noise term $\varepsilon_f$ to the spectral density:
            \begin{align}
                Y_f = S_f( \theta) + \varepsilon_f
                \label{eq:functional-likelihood}
            \end{align}
            $\varepsilon_f$ is assumed to follow a multivariate Gaussian distribution with zero mean and covariance $C_f(\eta)$, where $\eta$ is a precision parameter determining the scale of the noise: 
            \begin{align}
                \varepsilon_f \sim \mathcal{N}\left(0, C_f(\eta)\right)
                \label{eq:noise-distr}
            \end{align}
            This noise is generic and does not impose dependence between frequencies. One can consider the Gaussian form of noise as an approximation for non-central chi-distributed spectral noise that would arise from additive Gaussian noise in the temporal domain. The covariance matrix $C_f$ is defined as follows: 
            \begin{align}
                C_f(\eta) = \eta^{-1} R_f
            \end{align}
            where $R_f$ is the correlation matrix computed from the cross-spectral density of the signal, following \cite{camba2005estimating, friston2012dcm}. 
            
            Together, Eq.~\eqref{eq:functional-likelihood} and \eqref{eq:noise-distr} prescribe a spectral likelihood model that gives the probability of the spectrum $Y_f$ from parameters $\bm{\theta}:=\{\theta, \eta\}$: 
            \begin{align}
                p(Y_f\;|\;\bm{\theta}) = \mathcal{N}\left(Y_f\; ;\; S_f(\theta), C_f( \eta)\right)
                \label{eq:likelihood}
            \end{align}
            For most parameters, some values are more probable that others. This can be incorporated in the model through parameter priors.

        \subsection{Priors and constraints}
            Parameter priors reflect any knowledge we have about parameters before observing the data, e.g. preferred values and constraints. The choice of priors is delicate as it can make the model more complex, computationally intensive, or numerically unstable. To address these issues, we restrict priors to the well-behaved multivariate Gaussian family:
            \begin{align}
                p(\theta) &= \mathcal{N}(\theta; \mu_\theta, \Sigma_\theta)\\ 
                p(\eta) &= \mathcal{N}(\eta; \mu_\eta, \Sigma_\eta)\\
                p(\bm{\theta}) &= p(\theta)p(\eta) = \mathcal{N}(\bm{\theta}; \bm{\mu}, \bm{\Sigma})
            \end{align} The last equation reflects the standard assumption of independence between parameters of interest and precision parameters that govern noise. In brief, using Gaussian priors amounts to saying that we are only interested in the mean and (co-)variance of our model parameters. 

            Naturally, some model parameters do not fit the Gaussian assumptions, as they must respect some constraints --- for instance, a peak height must be positive. For this reason, model quantities in Eq.~\eqref{eq:periodic} and \eqref{eq:aperiodic} are mapped from parameters through link functions. Table~\ref{tab:link-functions} summarises link functions and prior expectations and covariances used in \bsd. The most common transform is the exponential transform, which gives a model parameter a log-normal distribution and ensures its positivity. 
            \begin{table*}[]
                \renewcommand{\arraystretch}{1.5}
                \centering
                \begin{tabular}{@{}cccc@{}}
                    \hline
                     Parameter & Symbol & Link function & Priors  \\
                    \hline\hline
                     Aperiodic exponent & $\alpha$ &  $\alpha(\theta)  = \exp(\theta_\alpha)$ & 
                     $\theta_\alpha \sim \mathcal{N}(0,3)$ \\
                     Aperiodic power & $\beta$ & $\beta(\theta)  = \exp(\theta_\beta)$ & 
                     $\theta_\beta \sim \mathcal{N}(0,3)$  \\
                     \hline
                     Peak power & $h_i$ & $h_i(\theta)  = \exp(\theta_{h_i})$ &  $\theta_{h_i} \sim \mathcal{N}(0,2)$ \\
                     Peak frequency & $m_i$ & $ m_i(\theta)  = L_i + \sigma({\theta_{m_i}})(U_i - L_i)$  &  $\theta_{m_i} \sim \mathcal{N}(0,2)$ 
                     \\
                     Peak width & $s_i$ & $ s_i (\theta) = \exp(\theta_{s_i})$&  $\theta_{s_i} \sim \mathcal{N}(0,2)$
                     \\
                     \hline
                \end{tabular}
                \caption{Parameters  of interest, their link function, and default priors. The function $\sigma$ appearing in the link function for the peak frequency is the hyperbolic arc-tangent. }
                \label{tab:link-functions}
            \end{table*}

            \begin{figure}[t!]
                \centering
                \begin{subfigure}{0.45\textwidth}
                    \centering
                    \caption{Link function for the peak frequency}
                    \includegraphics[width=1\columnwidth]{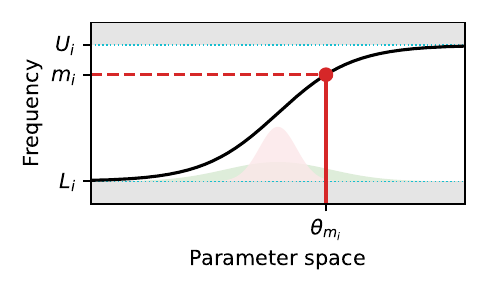}
                \end{subfigure}
                \begin{subfigure}{0.45\textwidth}
                    \caption{Peak in the frequency domain}
                    \centering
                    \includegraphics[width=1\columnwidth]{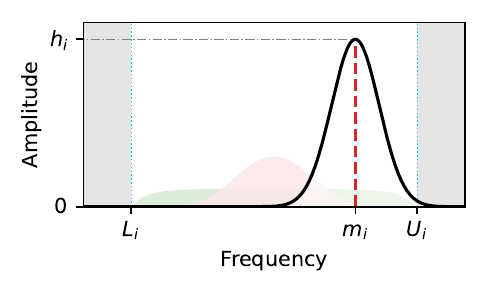}
                \end{subfigure}
                \caption{Parameterisation of the peak frequency $m_i$. In panel~(a), the unconstrained parameter $\theta_{m_i}$ is shown on the horizontal axis. Before it is used within the model, it is transformed to the peak frequency $m_i$ on the vertical axis (black curved line). Importantly, we see that the peak frequency is softly clipped between the band parameters $L_i$ and $U_i$ (dotted cyan lines). The resulting value of  $m_i$ (dashed red line) parameterises the mode of the Gaussian peak and is constrained between the frequency band limits, as shown in panel~(b). Green and pink shaded areas, on the x axis, display how Gaussian distributions over  $\theta_{m_i}$ with different variances (panel~(a)) induces different distributions over the frequency band (right~(b)). By default, we set priors as displayed in the green shaded area.}
                \label{fig:peak-parameterisation}
            \end{figure}
            Importantly, we have used an hyperbolic arc-tangent function to map each periodic component to disjoint sections of the frequency domain (referred to as \textit{frequency bands} in the following)  (Fig.~\ref{fig:peak-parameterisation}). Inherently, this means that \bsd{} is modelling a unique peak per frequency band. This restriction prevents the recruitment of several periodic components to fit a single peak. In addition, it ensures that the peaks estimated for different subjects are all in comparable frequency ranges, such that it makes sense to analyse them together. In other words, restricting the model to one peak per frequency band makes the model more interpretable and simpler to compare across subjects. 

        \subsection{Model estimation}
            As \bsd{} is a Bayesian method, model estimation should produce two quantities: the \textit{posterior distribution} -- i.e., the probability distribution of the parameters given the observed spectrum, $p(\bm{\theta}|Y_f)$ -- and the \textit{log marginal evidence} -- i.e., the log-probability of the observed spectrum under the model, $\ln p(Y_f)$. As this problem is generally intractable, \bsd{} relies on variational inference. Variational inference introduces an \textit{approximate posterior}, $q_\phi(\bm{\theta})$, and an approximate log marginal evidence: the \textit{free-energy}, $F$ (details about the algorithm, with accompanying code, can be found in \citet{zeidman2023primer}). At convergence, we have
            \begin{align}
                q_\phi(\bm{\theta}) &\approx p(\bm{\theta}|Y_f)\label{eq:post-approx} \\
                F &\approx \ln p(Y_f) \label{eq:vfe-approx}                 
            \end{align}
            which provides an approximate solution to the Bayesian inference problem. The probabilistic estimates of the parameters, $q_\phi(\bm{\theta})$, are then collated across subjects to enable     level analysis.

        \subsection{Group-level analysis}
            Once the spectral model of each subject as been estimated, one can conduct group-level analysis on model parameters. The motivation is to study and compare spectral effects at the group level, for instance, to quantify changes in the location or width of a peak  between pathological and healthy subjects. For this, \bsd{} leverages the Parametric Empirical Bayes (PEB) framework \cite{friston2016bayesian, zeidman2019guide2}. In brief, PEB allows one to rapidly perform Bayesian second-level analysis using a General Linear Model (GLM). Consider a group of $N$ individuals, where subject $k$ has spectral data $Y_k$. PEB models the data as: 
            \begin{align}
                Y_k &= S_f(\theta^{(1)}_k) + \varepsilon^{(1)}_k\\ 
                \theta^{(1)} &= \mathbf{X}^{(2)} \theta^{(2)}  + \varepsilon^{(2)} 
            \end{align}
             The first line is the spectral likelihood seen before, which accounts for within-subject effects: each subject is modelled with their own parameters $\theta^{(1)}_k$ and observation noise $\varepsilon^{(1)}_k$. Group-level effects are modelled by a GLM with design matrix $\mathbf{X}^{(2)}$ mapping parameters $\theta^{(2)}$ to subject-level model parameters. Between-subject variability is accounted for by the second-level noise parameter $\varepsilon^{(2)}$. Both the priors on the parameters governing the second-level noise and the parameter priors are assumed to follow multivariate normal distributions. In practice, \bsd{} uses the default priors provided in PEB routines in the Statistical Parametric Mapping (SPM) software, thus only the design matrix needs to be specified to perform a second-level analysis \cite{friston2016bayesian,zeidman2019guide2,zeidman2023primer}. Group-level analysis with PEB is conducted in three steps. First, one inverts each subject's model using gradient ascent on the variational free-energy to obtain the posterior distribution of each subject's parameters. Second, one specifies the second-level design matrix. Third, second-level model parameters are optimised to maximize the free-energy for the entire hierarchical model (accounting for both within- and between-subjects effects) \cite{friston2016bayesian}.

        \subsection{Bayesian model comparison}
            Bayesian model comparison is the Bayesian approach to making modelling decisions and mitigating between alternative explanations for the data \cite{stephan2009bayesian,friston2016bayesian,kass1995bayes}. In \bsd{}, it enables comparing different spectral models and making a principled selection, e.g., the number of periodic components to include in the model or their frequency band. 

             Consider two competing hypotheses $H_1$ and $H_2$, e.g., whether or not the data contains an alpha peak. These translate into two models $m_1$ and $m_2$ with different spectral likelihood and priors. Having observed spectral data $Y_f$, we want to assess whether one hypothesis is more probable than the other. Formally, $H_1$ being $\beta_{12}$ times more probable than $H_2$ after observing $Y_f$, translates as $p(m_1|Y_f) = \beta_{12} \, p(m_2|Y_f)$.  Assuming that both hypotheses are equiprobable a-priori, i.e. $p(m_1) = p(m_2)$,  Bayes rule gives
            \begin{align}
                \beta_{12} = \frac{p(m_1|Y_f)}{p(m_2|Y_f)} = \frac{p(Y_f|m_1) p(m_1)}{p(Y_f|m_2)p(m_2)} = \frac{p(Y_f|m_1)}{p(Y_f|m_2)} 
                \label{eq:bayes-factor}
            \end{align}
            The factor $\beta_{12}$ is called the Bayes factor \cite{kass1995bayes}. From Eq.~\eqref{eq:bayes-factor}, the Bayes factor between two equiprobable hypotheses is the ratio of the model evidence of their corresponding probabilistic models. Or more simply, selecting between hypotheses amounts to constructing their probabilistic models and comparing their ability to predict the data. A commonly accepted interpretation of the values of Bayes factors are shown in Table~\ref{tab:bayes_factor}. 
            \begin{table}[h!]
                \centering
                \begin{tabular}{@{}lll@{}}
                \toprule
                $\ln \beta_{12}$ & $\beta_{12}$ & Evidence towards $m_1$  \\ \midrule
                $< 0$& $< 1$ &  Negative (towards $m_2$) \\
                0 to 1 & 1 to 3 & Weak \\
                1 to 3 & 3 to 20& Positive \\
                3 to 5 & 20 to 150&  Strong \\
                $> 5$ & $> 150$ &  Very strong \\
                 \bottomrule
                \end{tabular}                
                \caption{Commonly accepted values of significance for Bayes factors, adapted from \citet{kass1995bayes}.}
                \label{tab:bayes_factor}
            \end{table}

            In variational inference, the marginal evidence is approximated by the free-energy (Eq.~\eqref{eq:vfe-approx}). Accordingly, we can estimate the log Bayes factor between the two models as: 
            \begin{align}
                \ln \beta_{12} \approx F_{m_1} - F_{m_2}
                \label{eq:log-bf}
            \end{align}
            where $F_{m_i}$ is the free-energy of the model after convergence. Thus, model comparison is conducted by first estimating each model and then comparing their free-energy.  
            
            Notably, Bayes factors are likelihood ratios - i.e., ratios of data likelihood under competing models. By the Neyman-Pearson lemma, mitigating between hypothesis using a likelihood ratio is the most powerful (i.e., sensitive) test across all possible levels of specificity. In other words, using Bayes factors as a decision criterion for model comparison maximises the Area Under the Curve (AUC) of the Receiver Operating Characteristic (ROC) curve. For \bsd{}, this ensures that Bayesian model comparison is the most powerful way to detect periodic components. 

        \subsection{Additions to the spectral model}
            The proposed generative model for the spectrum can be easily expanded without changing the procedure detailed above. Extensions include modelling the effect of filters, condition-specific effects, and source level data.  
            \subsubsection{Filter modelling}
                In practice, the spectral data has undergone several filtering steps before being analysed. In \bsd{}, the knowledge of the filtering history and each filter's parameter can be leveraged to account for information lost in filtering. Having applied $K$ filters, where filter $k$ has transfer function $H_k$, the spectral model becomes 
                \begin{align}
                    Y_f = \prod_{k = 1}^K |H_k(f)| S_f(\theta) + \varepsilon_f
                \end{align}
                This is especially important as the covariance of $\varepsilon_f$ is estimated from the data. Thus, filtered regions will invariably have a low covariance. The model might thus become over-precise in filtered regions. Including the filtering effect on the spectral amplitude acknowledges this reduction in variance, and reduces the risk of overfitting data in filtered regions.  
                
            \subsubsection{Condition-specific effects}
                In most empirical studies, the interesting results do not lie in the absolute characterisation of data features but rather in quantifying their relative change between different experimental conditions --- for instance, quantifying how a much the peak height varies between rest and task. \bsd{} accounts for condition-specific effects at the individual level using a linear mapping from conditions to parameters. Assuming several conditions, where $Y_{f,c}$ is the spectral data for condition $c$, Eq.~\eqref{eq:functional-likelihood} is adapted as 
                \begin{align}
                p(Y_{f,c}\;|\;\bm{\theta}) = \mathcal{N}\left(Y_{f,c}\; ;\; S_f(X_c \,\bm{\theta}), C_f( \eta)\right)
                \label{eq:cond-effect}
                \end{align}
                $X_c$ plays the role of a design matrix and maps a linear combination of parameters to each condition. 

        \subsubsection{Source and sensor space}
            \bsd{} allows one to analyse the spectral data in the sensor or source domain. In the source domain, both the forward model and prior on source location must be specified. Then, a current source density approach is used to project the modelled spectrum at source level onto the measured spectrum at sensor level. Importantly, the source locations are formulated as priors, thus, the location of precise sources can be refined from the data. In both source and sensor space modes, the frequency and width of each periodic components can either be shared between all channels or estimated independently for each channel. In addition, the formulation allows one to estimate at the same time a common aperiodic component and several aperiodic components for each individual source or sensor. This can help distinguish region-specific aperiodic components from the global background activity.
        
\section{Results}
    \subsection{Face validity}
    \subsubsection{Simulation procedure}
        We first establish the face validity of \bsd{} on simulated data. We construct an example that features two conditions, labelled "rest" and "task". We prescribe a spectrum with a large peak in the alpha band (defined as 8 to 12~Hz) and a smaller peak in the beta band (defined as 12 to 30~Hz) in the rest condition. Akin to eyes close / eyes open experiments, the effect of the task condition is to severely attenuate both peaks. In addition, we introduce a continuous covariate, e.g. representing age, that differs for each subject and modulates their spectral parameters. In Figure~\ref{fig:sim-gen-process}, we display the average spectrum in both conditions as a function of the continuous regressor.  
        \input{figures/simulation-process}
        
        Having established  this generative process, we sample the spectrum of 32 artificial subjects. To sample a  spectrum from a subject, we first sample a random value of the continuous covariate, compute the corresponding spectral parameters, and add noise to these parameters. We then compute the spectrum for frequency between 1 and 32~Hz and add observation noise. This gives 64 spectra, shown in Fig.~\ref{fig:sim-gen-process}, which play the role of our experimental observations. 
        
    \subsubsection{Model selection}
        From the observed spectrum, we first need to select a model for our data. The model space consists of all models to be compared, where each model is defined by a particular set of frequency bands. Without any prior knowledge on the frequency content of the signal, one can populate the model space by exploring all possible combinations of a set of frequency bands of interest. In practice, one can refine the model space by introducing some hypotheses about what periodic peaks might be present. In this example, a visual inspection of the individual spectra and their median split suggests the presence of an alpha peak and to a lesser extend, that of a beta peak. The next step is to confirm the presence of each peak using Bayesian model comparison. In this example, there are two possible peaks and thus four candidate models: no peaks, beta peak only, alpha peak only,  or both peaks. This forms the model space, which is illustrated in Figure~\ref{fig:sim-model-space}. 
        \input{figures/simulation-model-comparison}

        We next invert each model in our model space on data from each artificial subject. After optimisation, we obtain an estimate of the marginal evidence (free energy) under each model and each subject (Fig.~\ref{fig:sim-model-space}~(a)). We observe that models that include an alpha peak have much  higher evidence than models without it (Fig.~\ref{fig:sim-model-space}~(a)). For the beta peak, the difference is more subtle. We average the evidence with and without an alpha peak, and that with and without a beta peak. Results are shown in Fig.~\ref{fig:sim-model-space}~(b). We see that the presence of both alpha and beta peak increases the free-energy. To confirm the significance of these findings, we compute the free-energy difference, which is a proxy for the log Bayes factor (Eq.~\eqref{eq:log-bf}). We observe a log Bayes factor of more that 100 for the alpha peak, and more than 6 for the beta peak. Thus, according to the interpretation of log Bayes factors~(\ref{tab:bayes_factor}), we can conclude that we have found very strong evidence for the presence of an alpha peak, and strong evidence for the beta peak. The best model for our data is model 4, which includes both alpha and beta peaks. Indeed, this corresponds to the model used to generate the data, and highlights the adequacy of Bayesian Model Comparison to uncover the right model. 
        
    \subsubsection{Effect of the continuous regressor}
        \input{figures/simulation-peb}
        So far, we have identified that the model with both alpha and beta peaks is the best model for this data. For this, we have inverted a model for each subject, and obtained both the model evidence and the posterior parameter distribution. In addition, for each subject, we have the value of the continuous covariate, e.g.  age. We now want to test whether the within-subject parameters are associated with the continuous covariate at the group level. This translates into writing a general linear model that explains the estimated posterior distribution of the subjects' parameters as a function of the continuous covariate, plus an intercept capturing the  average parameter value for the population. The design matrix corresponding to this linear model is displayed in Figure~\ref{fig:sim-peb-design}. 

        We use PEB to estimate the model coefficients. After applying PEB, for each of the 8 parameters of our model (2 for the aperiodic component, and $2\times 3$ for the periodic component), we have the estimated parameter average across the population, and its association  with the covariate. These two parameters configure the intercept and slope of a linear mapping of the parameter against the covariate. As we are working in a Bayesian framework, we also have confidence intervals for these two effects, and thus for the entire curve (in Bayesian analysis these are  referred to as credible intervals). In other words, we can recover the mean and 95\% interval for the power and exponent of the aperiodic component (Fig.~\ref{fig:sim-peb-aperiodic}) and the power, frequency, and width of the alpha (Fig.~\ref{fig:sim-peb-alpha}) and beta (Fig.~\ref{fig:sim-peb-beta}) modes. We can see that for most quantities, the true effect falls within the 95\% confidence interval of our model. This is not the case for the frequency decrease of the alpha mode frequency, where the estimated effect is overconfident. Overconfidence in variational inference is a known effect that has been reported elsewhere (see \cite{friston2002classical,daunizeau2009variational}). 
        
        Because we have the full model of how parameters change with the covariate, we are now in position to interpolate the spectrum over the entire covariate range of the population. This gives the plot  Figure~\ref{fig:sim-peb} (c), which we can compare with that of the generative process (Figure~\ref{fig:sim-gen-process}). We see that \bsd{} recovers all the effects of interest. 
        
    \subsection{Sensitivity and specificity: comparison with \fooof{}}
        \input{figures/simulation-snrs}
        In this section, we evaluate the sensitivity and specificity of \bsd{} when used to assess the presence of a peak in a spectrum. We also compare both metrics with an established method, \fooof{}. In particular, we are interested in observing how the statistical power of both methods changes with the peak height, the peak location, and the amount of observation noise. 

            We generate spectra with a single peak within a frequency band. We investigate the following frequency bands:  delta (1 to 4~Hz), theta (4 to 8~Hz), alpha (8 to 12~Hz), beta (12 to 30~Hz), and gamma (30 to 64~Hz). For each band, we generate 1,024 random spectra over a frequency grid from 1 to 64~Hz with 0.5 Hz resolution. Each spectrum has a random peak location and width, and random aperiodic parameters. The peak height is set to a fixed SNR, defined as a fixed factor of the sum of the aperiodic component and variance of the output noise. Similarly, the variance of the output noise is set to a fixed fraction of the scale of the aperiodic component. 
        
        For each spectrum, we invert two \bsd{} models: a null model, without any peak, and an informed model, with a peak in the correct frequency band. After inversion, we compute the log Bayes factor between the two models and use it as a criterion to determine whether a peak is present or absent. Separately, we estimate a \fooof{} model using the \texttt{specparam} Python package with default parameters (from \href{https://github.com/fooof-tools/fooof/tree/v1.1.0}{\texttt{specparam~V1.1.0}}). To determine whether a correct peak is detected by \fooof{}, we identify the estimated peak with the closest mean frequency to the  peak used to simulate the data, and use its estimated height as a decision criterion.

        Having defined a criterion for each model, we compute its Receiver Operating Characteristics. This is done by plotting the specificity against the sensitivity of the model, under all possible values of decision threshold. In other words, we set a numerical threshold to compare the criterion against and determine the presence of a peak, compute the specificity and sensitivity of the test, and repeat these steps for all possible threshold values. Sample ROC curves are shown in Fig.~\ref{fig:sim-snrs}. The ROC can be summarised by the Area Under its Curve (AUC). We compute the AUC of the ROC for \bsd{} and \fooof{}. We then plot the AUC of the ROC as a function of the peak SNR. Key results are displayed in Fig.~\ref{fig:sim-snrs}, and complete results are attached in the supplementary materials. 
        
        For all frequency bands and SNRs, we see that both \bsd{} and \fooof{} manage to uncover the true peak when its height is large enough. Interestingly, for all plots, we see that \bsd{} is much better than \fooof{} in detecting peaks of intermediate height, and has a higher AUC overall. This is a nice illustration of Neyman-Pearson lemma, which states that likelihood ratios have the maximum ROC-AUC. 
        
       This result may be contrasted against very small peaks with SNR under -10~dB. There, \bsd{} can have a ROC-AUC comparable or under that of \fooof{} and even under 0.5 (see Supplementary materials). Naively, the latter case means that \bsd{} performs ``worse'' than deciding randomly whether a peak is present. This reflects a more subtle and important effect inherited from the use of variational inference. Indeed, we use the free-energy to compute the log Bayes factor. By construction, the free-energy expresses a trade-off between accuracy and complexity. When a peak is too small,  models with and without that peak have the same accuracy but introducing a peak increases the model complexity. Thus, the approximate log Bayes factor points us towards the simplest explanation for the data, i.e., the model without a peak, which does not correspond to the generative process and thus counts as a false negative. Hence, this result is observed only for very low SNRs (less than -10~dB, i.e., a noise amplitude 10 times larger than the signal amplitude) and reflects a conservative aspect of \bsd{}, which favors a simpler explanation (i.e., not reporting a peak) when the data does not justify a more complex one (i.e., a peak that can be mistaken for noise).

\subsection{Worked example: applying \bsd{} to the LEMON dataset}
    \subsubsection{Data presentation}  
        In this section, we show how \bsd{} can be deployed to analyse the frequency content of real EEG data and the variation of spectral parameters with age. For this, we use the LEMON dataset \cite{babayan2019mind, mendes2019functional}, which features 204 subjects and includes experimental EEG data with eyes closed and eyes open conditions. In addition, the age range of each subjects is reported, with a resolution of 5 years. 

        We aim at building a simple model of the spectral density at channel Oz in both conditions, expressed using a commonality and condition-specific parameterisation (Eq.~\eqref{eq:cond-effect}). In addition, we leverage the statistical power of the sample size to estimate a linear model of parameters as a function of age, across the entire group.  
        
    \subsubsection{Model selection}
        \input{figures/lemon-bmc}
        In the spectra for all subjects and in both conditions, a strong alpha peak is apparent, as well as possible delta, theta, and beta peaks. To assess the probability for the presence of these peaks, we invert the full model space for each of the subjects. After having estimated all models, we use Bayesian model comparison to identify the best model. Results of model comparison are shown in Fig.~\ref{fig:lemon-bmc}. We report a winning model having all three peaks: delta, alpha, and beta, with decisive evidence for theta and alpha and strong evidence for beta. The data does not support the inclusion of the delta peak.

    \subsubsection{Effect of age}
         \input{figures/lemon-overall}
        After identifying the optimal model, we construct a linear model that links the model parameters to subject age, similar to our simulated example. Since we only have age ranges rather than exact ages, we approximate the subject's age using the mean of each age range. Parameter estimation at the group level is achieved using PEB. In Fig.~\ref{fig:lemon-overall}, we display sample spectra for all subjects, along with averaged spectra post-median split by age. Additionally, we depict the interpolated group-level spectral model across the entire age spectrum. While the aging effect on the spectra from the median split is minimal, BSD effectively captures a prominent aging effect on the overall spectrum. This underscores the advantage of employing a fully Bayesian pipeline with PEB, which provides robust parameter estimates for the assumed generative model, enabling straightforward interpolation and rigorous spectral analysis. 
        
        \input{figures/lemon-modes}
        Results for individual components are shown in Fig.~\ref{fig:lemon-modes}. We see that the power of the aperiodic component exhibits a slight decrease with age, and the slope of the aperiodic component decreases (the exponent increases) with age (Fig.~\ref{fig:lemon-aperiodic}). We report a decrease of theta frequency, with an increase of the power and a decrease of the width (Fig.~\ref{fig:lemon-theta}). As expected, we find a strong decrease of the alpha peak power and  frequency with age, as well as an increase of the peak width (Fig.~\ref{fig:lemon-alpha}). Surprisingly, we report an increase of the power of the beta peak with age, together with a decrease in frequency and an increase in width (Fig.~\ref{fig:lemon-beta}). We report no effect of age on the between-condition effects.

        Most of these finding are in line with the existing aging literature: age is related to an overall slowing of brain waves and a decrease of the slope of the aperiodic component \cite{trondle2023decomposing, merkin2023age, hill2022periodic, thuwal2021aperiodic}. Here, \bsd{} allows us to quantify each slope, as well as their confidence intervals, in a simple but rigorous manner.


\section{Discussion}
    Our study introduces \bsd{}, a novel method for analyzing neural spectral power. Through a series of simulations and application to real EEG data, we have demonstrated its utility in identifying spectral peaks, selecting appropriate models, and investigating relationships between spectral parameters and continuous covariates such as age. Here, we discuss the implications of our findings and the potential future directions for research in this area.
    
    Our results from the simulation study showcase the efficacy of \bsd{} in model selection and parameter estimation. By comparing different model configurations using Bayesian model comparison, we were able to accurately identify the presence of spectral peaks and select the most appropriate model for the data. Furthermore, our comparison with \fooof{} highlights the strengths of \bsd{} in detecting spectral peaks, particularly in scenarios with intermediate peak heights. The use of variational inference in \bsd{} allows for a principled approach to model selection, balancing between model accuracy and complexity. This makes \bsd{} a conservative method, which favors simpler models when the data provide insufficient support for the presence of very small peaks.
    
    One notable advantage of \bsd{} is its ability to incorporate continuous covariates into the analysis, as demonstrated using both simulated and real data. By modeling the relationship between spectral parameters and covariates using PEB, we were able to quantify how neural spectral features vary continuously with age. 

    The principled approach to group-level analysis using \bsd{} could enable new studies of large datasets in clinical neuroscience, for instance to understand the effects of aging and the development of dementia in relation to environmental factors \cite{scally2018resting}. In addition, \bsd{} could be leveraged to analyse experimentally-driven spectral changes reflecting slow mechanisms such as adaptation and learning, in conjunction with existing methodologies for analysing multiscale time series \cite{medrano2024linking}. Less trivially, \bsd{} could be used to analyse resting state fMRI signals, providing a phenomenological complement to existing mechanistic approaches \cite{friston2014dcm}.
    
    One promising direction for future research is the integration of \bsd{} with dynamic causal modeling (DCM) for a comprehensive analysis of neural spectra. DCM offers a mechanistic framework for understanding causal interactions between different brain regions and how they give rise to observed neural activity. We suggest that \bsd{} can be first be used to answer \textit{phenomenological} questions, in other words, summarise \textit{how} data features are changing. Then, DCM can be used to answer \textit{mechanistic} questions and understand \textit{why} data features are changing, relating changes in spectral power and coherence to changes in effective connectivity. Together, \bsd{} and DCM would allow one to automatically identify robust spectral features and to relate them to the underlying mechanisms at play. 
    
\section{Conclusion}
    In this work, we introduced \bsd{}, a Bayesian framework for constructing, analysing, and comparing spectral models of neural power. We have shown that the fundamentals of \bsd{} rest on well-established Bayesian methods that allow one to rigorously compare hypotheses and perform group-level statistics on spectral data. Our results on simulated data show that \bsd{} yields robust yet sensible results, outperforming \fooof{} in identifying spectral peaks. In addition, we have shown how \bsd{} can be used straightforwardly to perform group-analysis on large datasets. 
    
    In conclusion, \bsd{} offers a powerful and flexible approach for analyzing neural spectral power, with applications ranging from basic research on brain function to clinical studies of neurological and psychiatric disorders. By leveraging advanced Bayesian modelling techniques with expressive parametric models of neural spectra, \bsd{} opens up new avenues for investigating the complex dynamics of brain activity and their relationship to behavior and cognition.

\section*{Supporting Information}
    The Wellcome Centre for Human Neuroimaging is supported by core funding from Wellcome [203147/Z/16/Z]. NA and RAS are supported by a Wellcome Principal Research Fellowship to Eleanor Maguire [210567/Z/18/Z].
    
\section*{Code availability}
    BSD is made available as a toolbox as part of the SPM software package, available at \url{https://github.com/spm/spm}. A tutorial on \bsd{} is available on SPM documentation website at \url{https://www.fil.ion.ucl.ac.uk/spm/docs/}.
\section*{Data availability}
    The LEMON dataset used in the worked example is openly available at \url{https://fcon_1000.projects.nitrc.org/indi/retro/MPI_LEMON.html}. 
\bibliographystyle{apalike}
\bibliography{references}


\appendix

\begin{sidewaysfigure*}
    \centering
    \includegraphics[width=\textwidth]{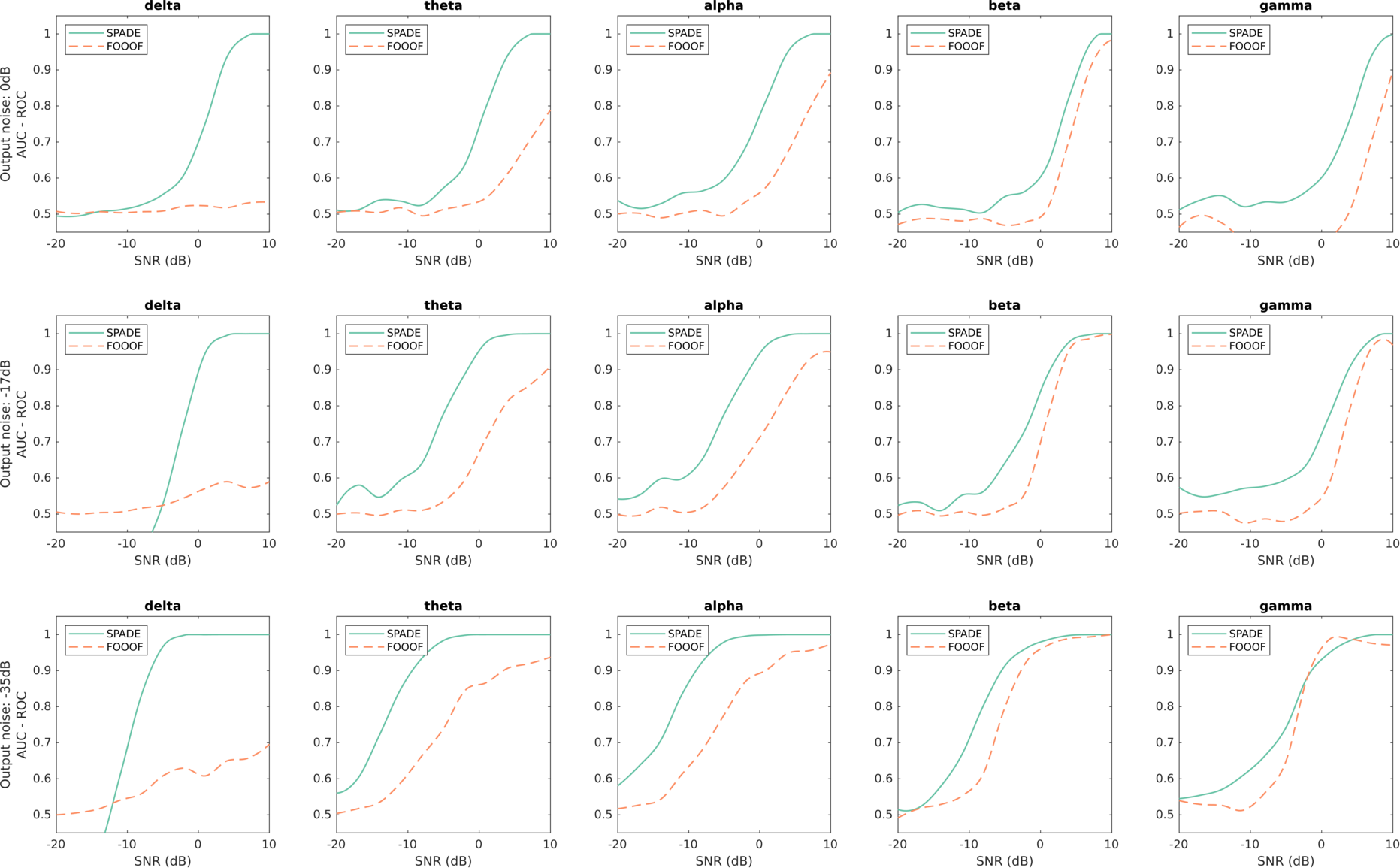}
    \caption*{Supplementary material 1: AUC-ROC, i.e. the detection score, as a function of the peak signal-to-noise ratio in decibels for both BSD (in plain green) and FOOOF (in dashed orange). Each row shows a different level of output noise:  $0$dB (top), $-17$dB (middle), and $-35$dB (bottom). Columns corresponds to the frequency band in which the peaks are located: delta (1-4~Hz), theta (4-8~Hz), alpha (8-12~Hz), beta (12-30~Hz), and gamma (30-100~Hz). For FOOOF, the default parameters of the official Python implementation (\texttt{specparam}) have been used.}
    \label{fig:snr-full}
\end{sidewaysfigure*}

\end{document}

%% file: figures/simulation-process.tex
\begin{figure*}[t!]
    \centering
    \begin{subfigure}[b]{0.49\textwidth}
        \centering
        \includegraphics[width=\textwidth]{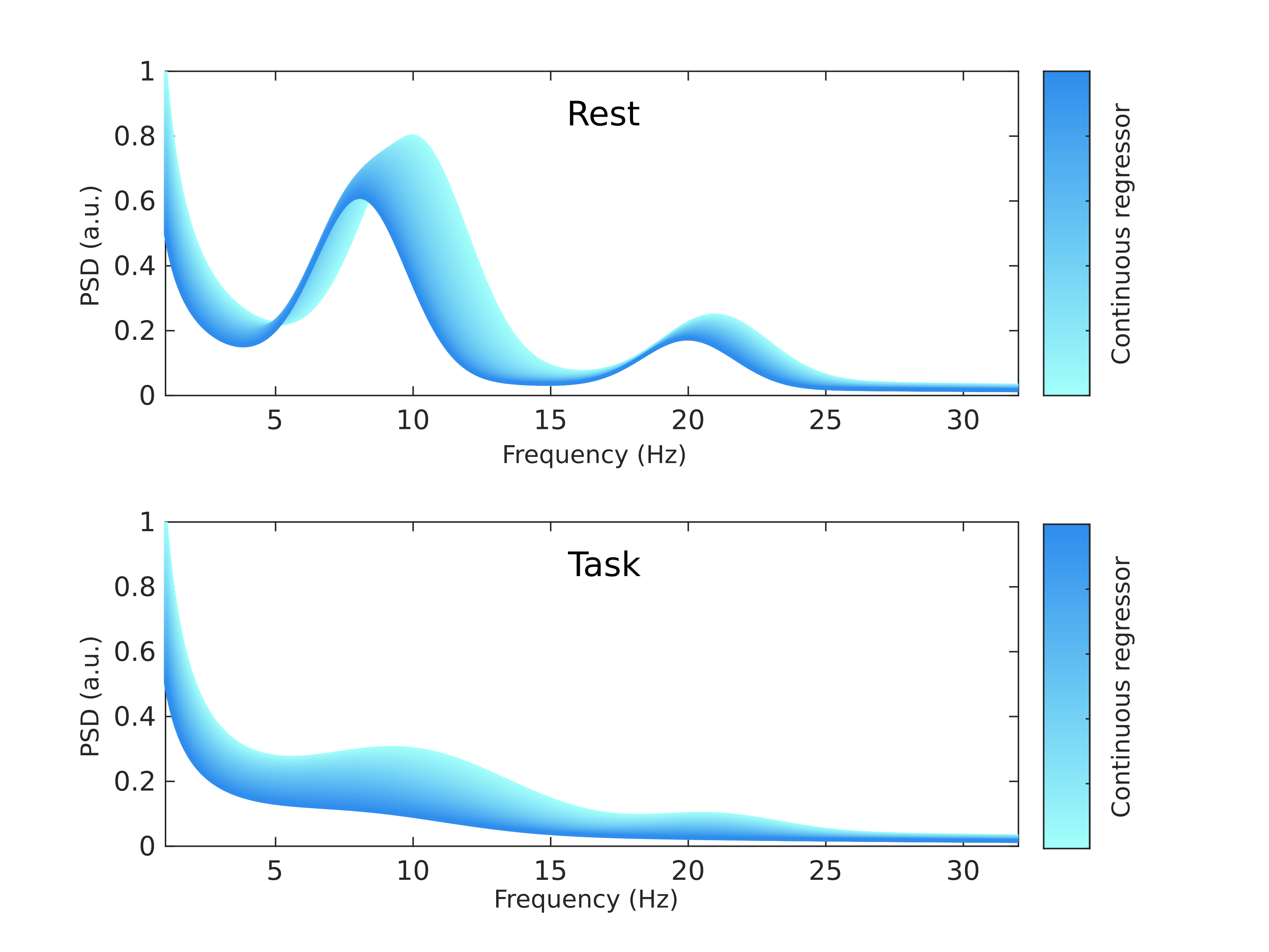}
        \caption{Generative process}
        \label{fig:sim-gen-process-a}
    \end{subfigure}
    \begin{subfigure}[b]{0.49\textwidth}
        \centering
        \includegraphics[width=\textwidth]{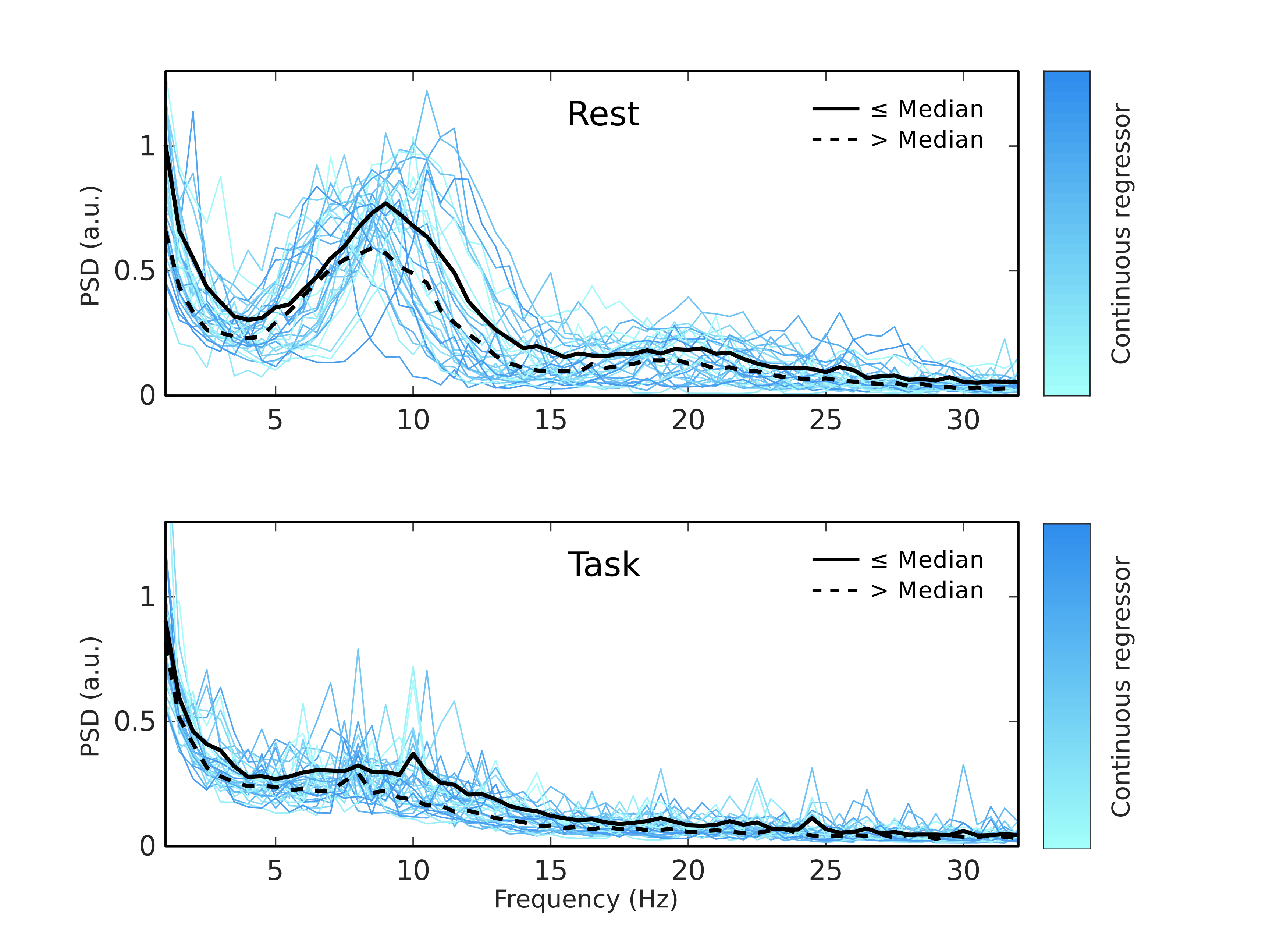}
        \caption{Simulated subjects spectrums}
        \label{fig:sim-gen-process-b}
    \end{subfigure}
    \caption{Spectrum and samples of the simulated example. The left panels show the generative process, i.e., the model used to generate the data, in the rest and task conditions. The spectrum in each condition is influenced by a continuous regressor, indicated by the shading of the line, which can be seen as a subject trait such as age. The right plot shows the sampled spectra in both conditions for the 32 subjects. The plain black line shows the spectral average for subjects with a continuous regressor value below the population median, and the dashed line corresponds to the average for subjects above the median.  }
    \label{fig:sim-gen-process}
\end{figure*}

%% file: figures/simulation-model-comparison.tex
\begin{figure*}[t!]
\centering
\begin{subfigure}[b]{0.26\textwidth}
    \centering
    \includegraphics[width=\textwidth]{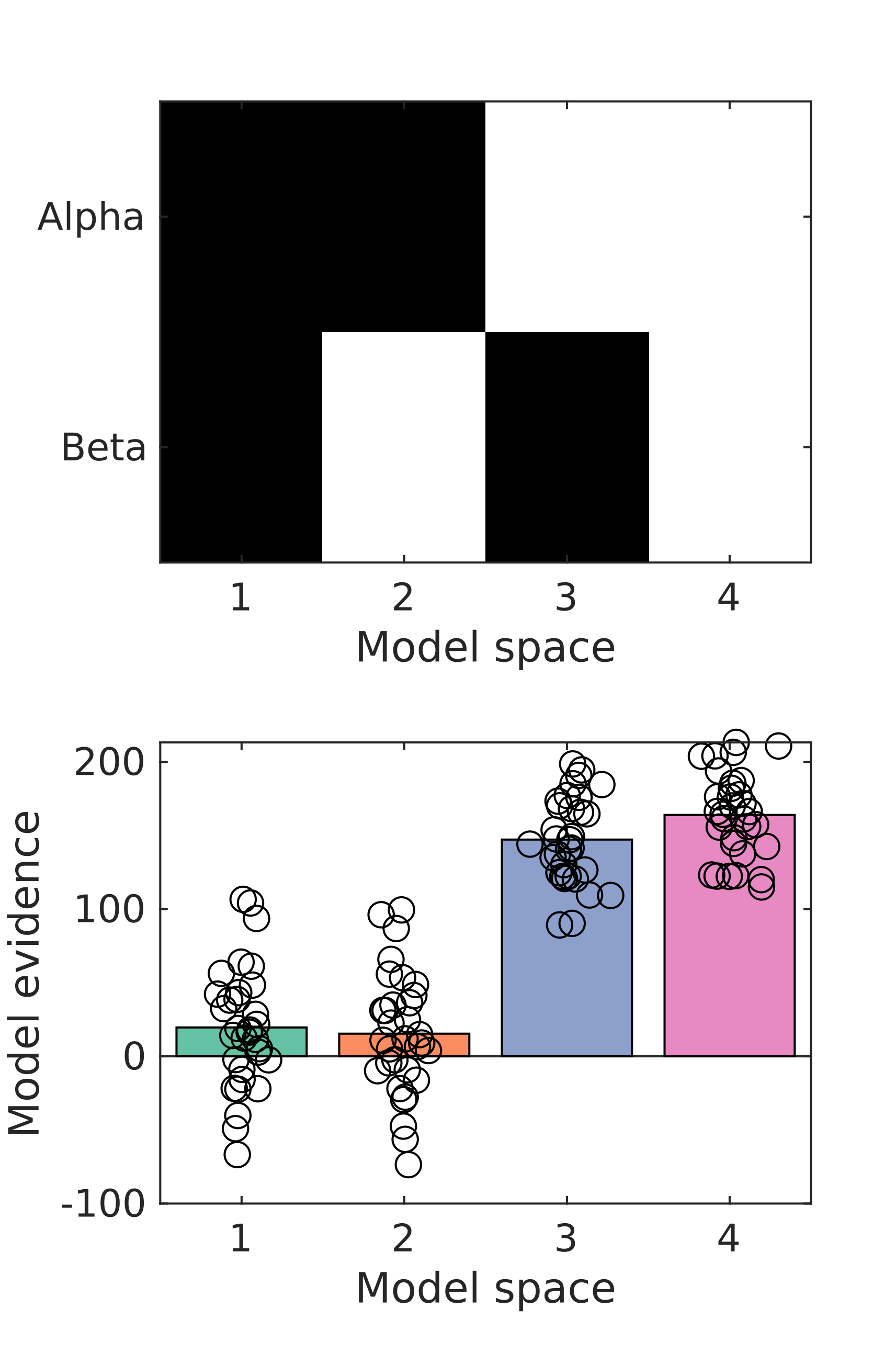}
    \caption{Model space and evidence}
    \label{fig:sim-model-space-full}
\end{subfigure}%
\begin{subfigure}[b]{0.73\textwidth}
    \centering
    \includegraphics[width=\textwidth]{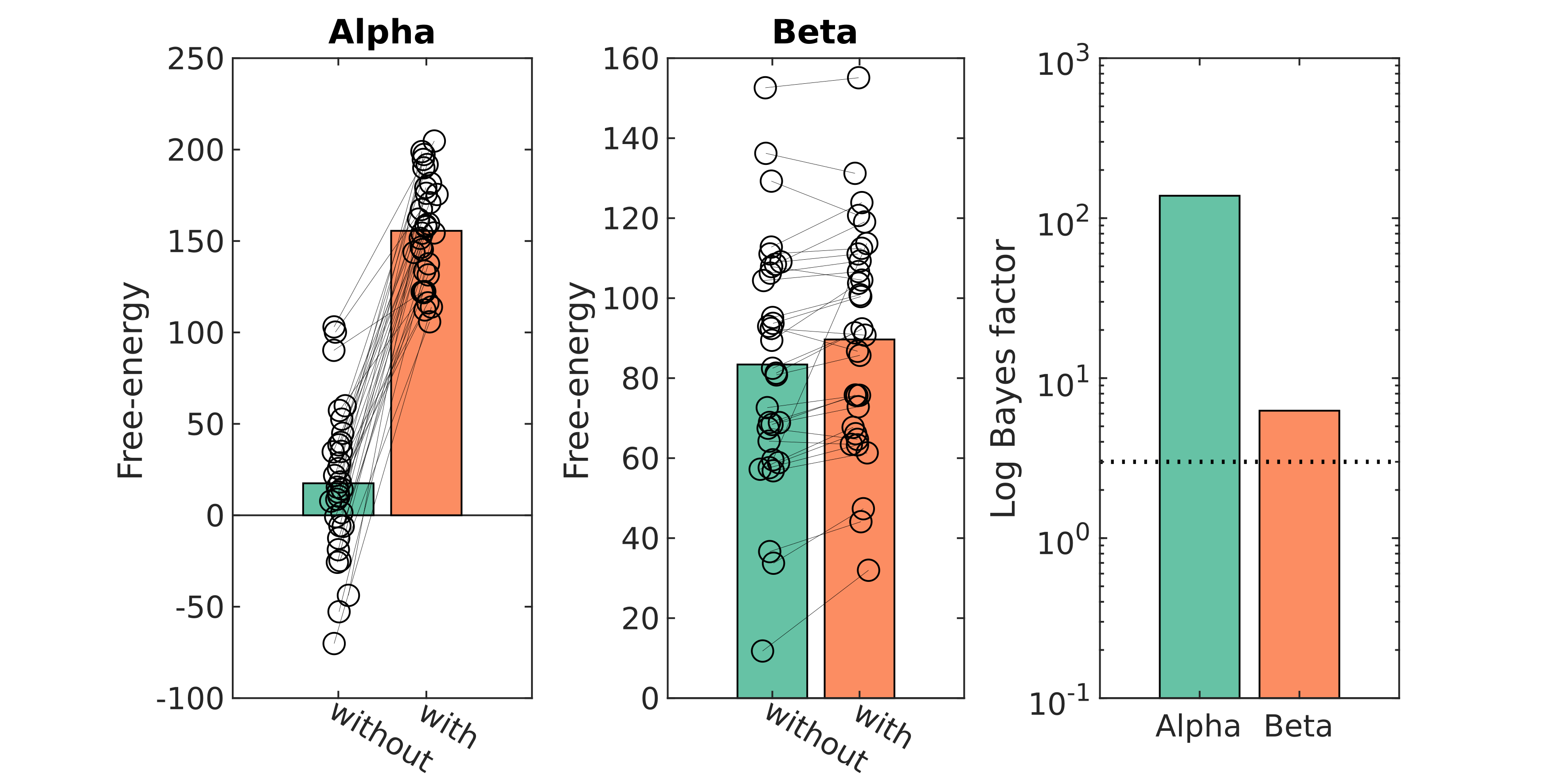}
    \caption{Pooled free-energy and log Bayes factor}
    \label{fig:sim-model-space-bf}
\end{subfigure}
\caption{Model comparison and frequency band selection. The model space is created by having alpha and beta peaks either absent or present, as shown on the upper left panel (black=absent, white=present). Models 3 and 4 have an alpha peak and models 2 and 4 have a beta peak. Model evidences, shown on lower left plot, are obtained by inverting each model for each subjects. In this panel, each marker represents the log-evidence of a subject, and the bars show the group average. The two plots on the left of panel b) are obtained by averaging the free-energy of models without and with a particular peak. The average free-energy increase caused by introducing the alpha or beta peak is the log-Bayes factor, shown on the right plot of panel b). The dotted horizontal lines highlight a log Bayes factor of 3, which corresponds 20 times probability factor and is commonly considered as strong evidence (see~\ref{tab:bayes_factor}).  }
\label{fig:sim-model-space}
\end{figure*}

%% file: figures/simulation-peb.tex
\begin{figure*}[t!]
    \centering
    \begin{subfigure}[t]{\textwidth}
    \begin{subfigure}[t]{0.09\textwidth}
        \centering
        \includegraphics[width=1\textwidth]{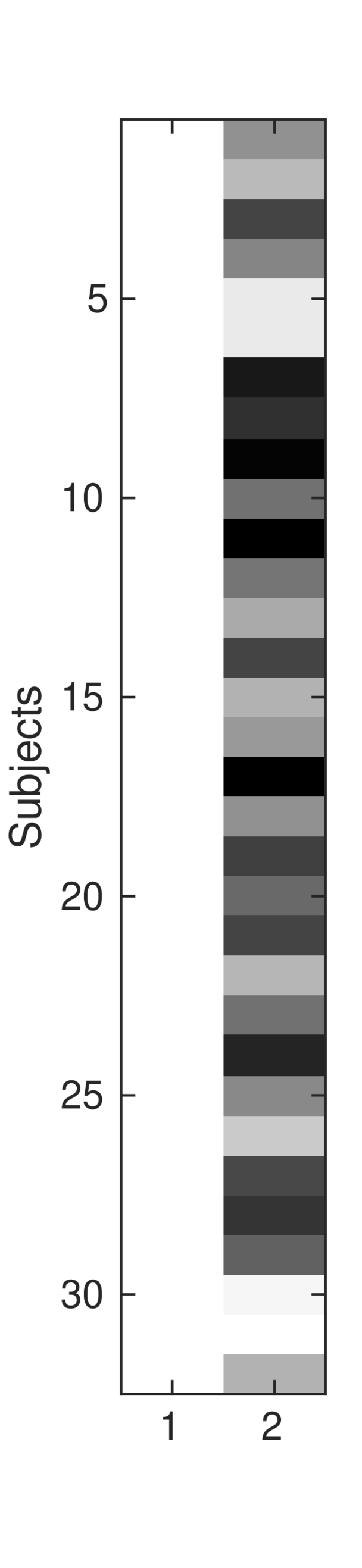}
        \caption{Design matrix}
        \label{fig:sim-peb-design}
    \end{subfigure}%
    \begin{subfigure}[t]{0.53\textwidth}
        \centering
        \includegraphics[width=1\textwidth]{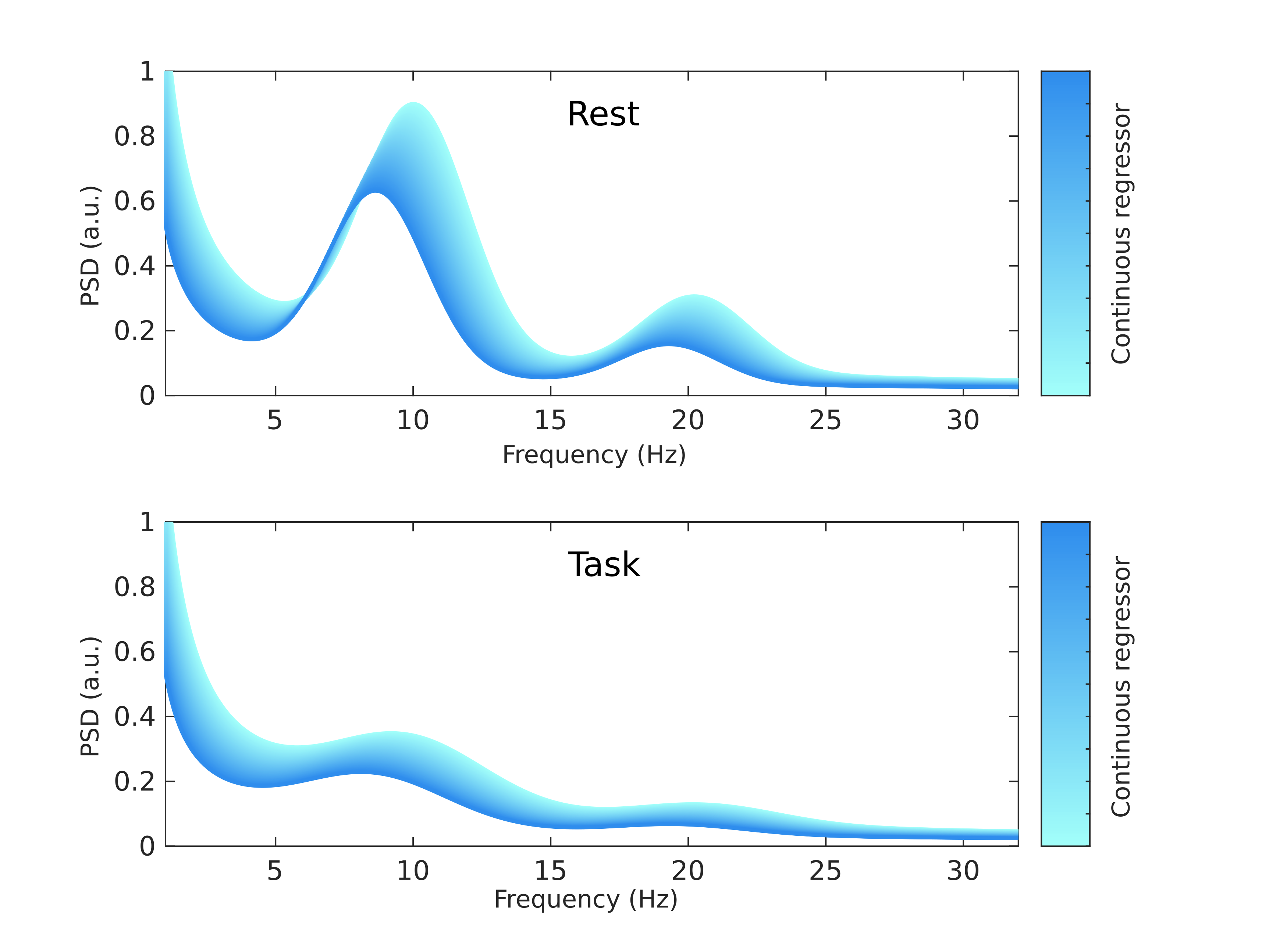}
        \caption{Recovered spectrum}
        \label{fig:sim-peb-recovered}
    \end{subfigure}%
    \begin{subfigure}[t]{0.31\textwidth}
        \centering
        \includegraphics[width=1\textwidth]{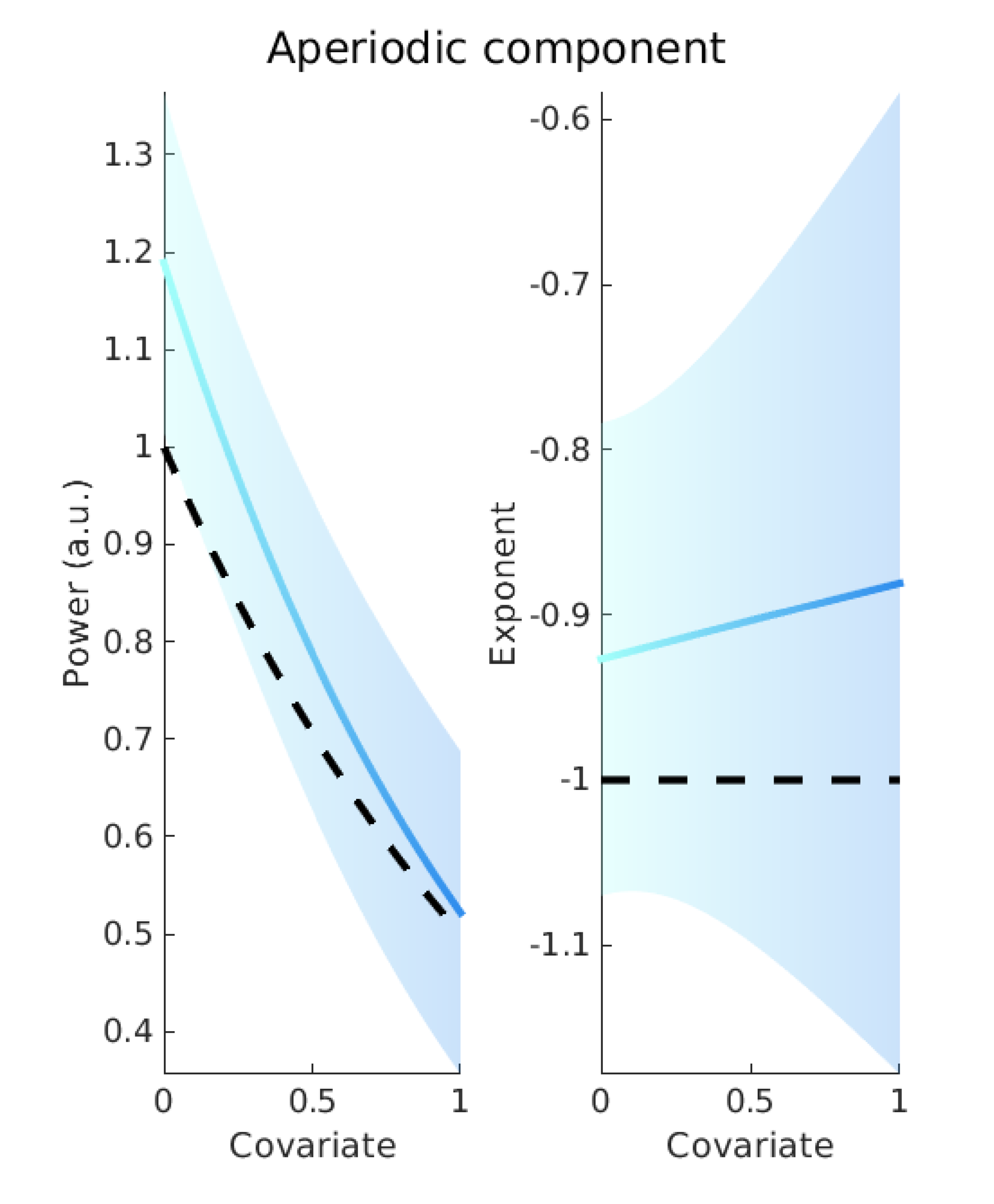}
        \caption{Aperiodic component during rest}
        \label{fig:sim-peb-aperiodic}
    \end{subfigure}
    \vspace{2em}
    \end{subfigure}
    \vspace{3em}
    \begin{subfigure}[b]{\textwidth}
    \begin{subfigure}[b]{0.49\textwidth}
        \centering
        \includegraphics[width=1\textwidth]{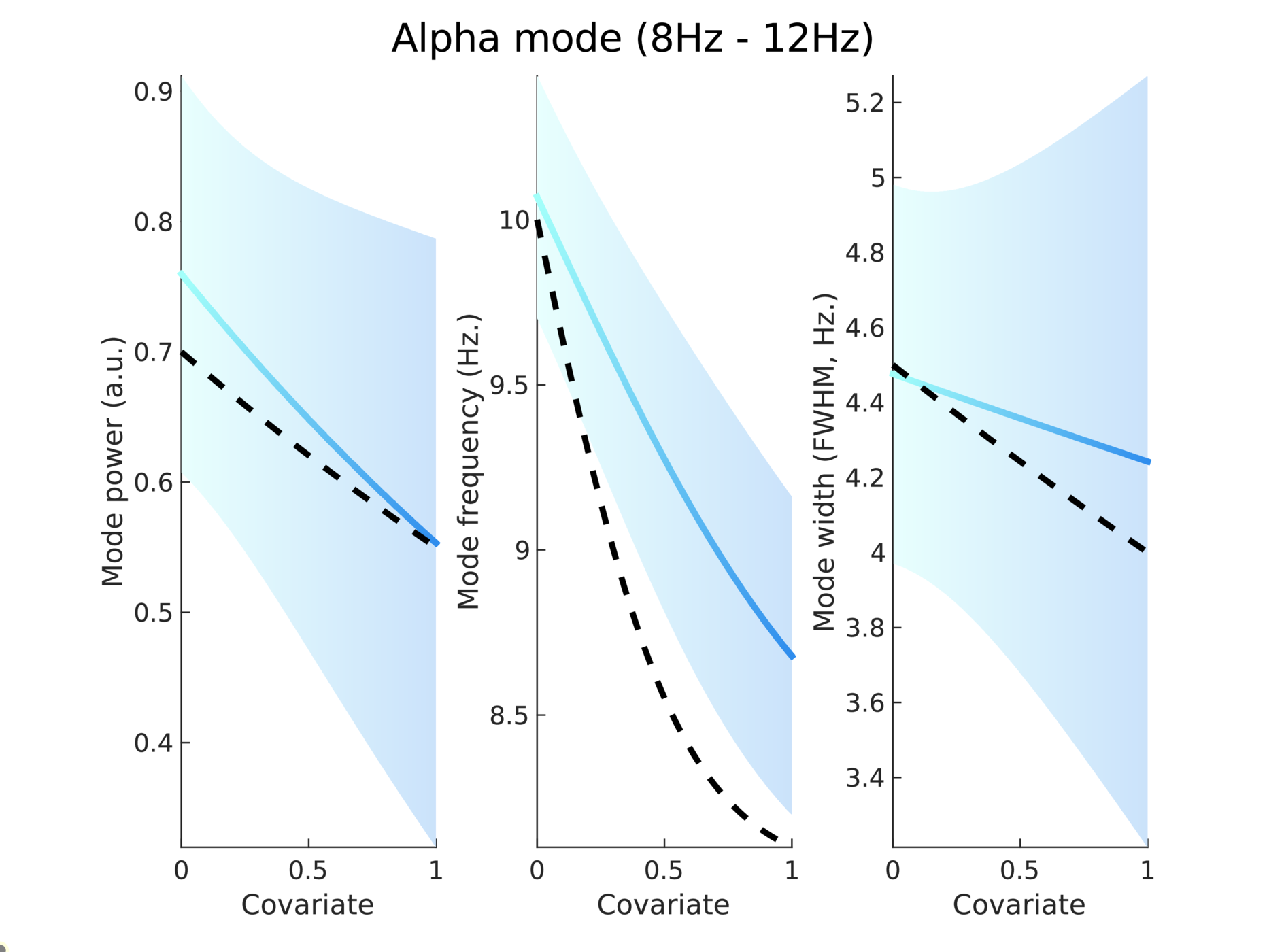}
        \caption{Alpha mode during rest}
        \label{fig:sim-peb-alpha}
    \end{subfigure}%
    \begin{subfigure}[b]{0.49\textwidth}
        \centering
        \includegraphics[width=\textwidth]{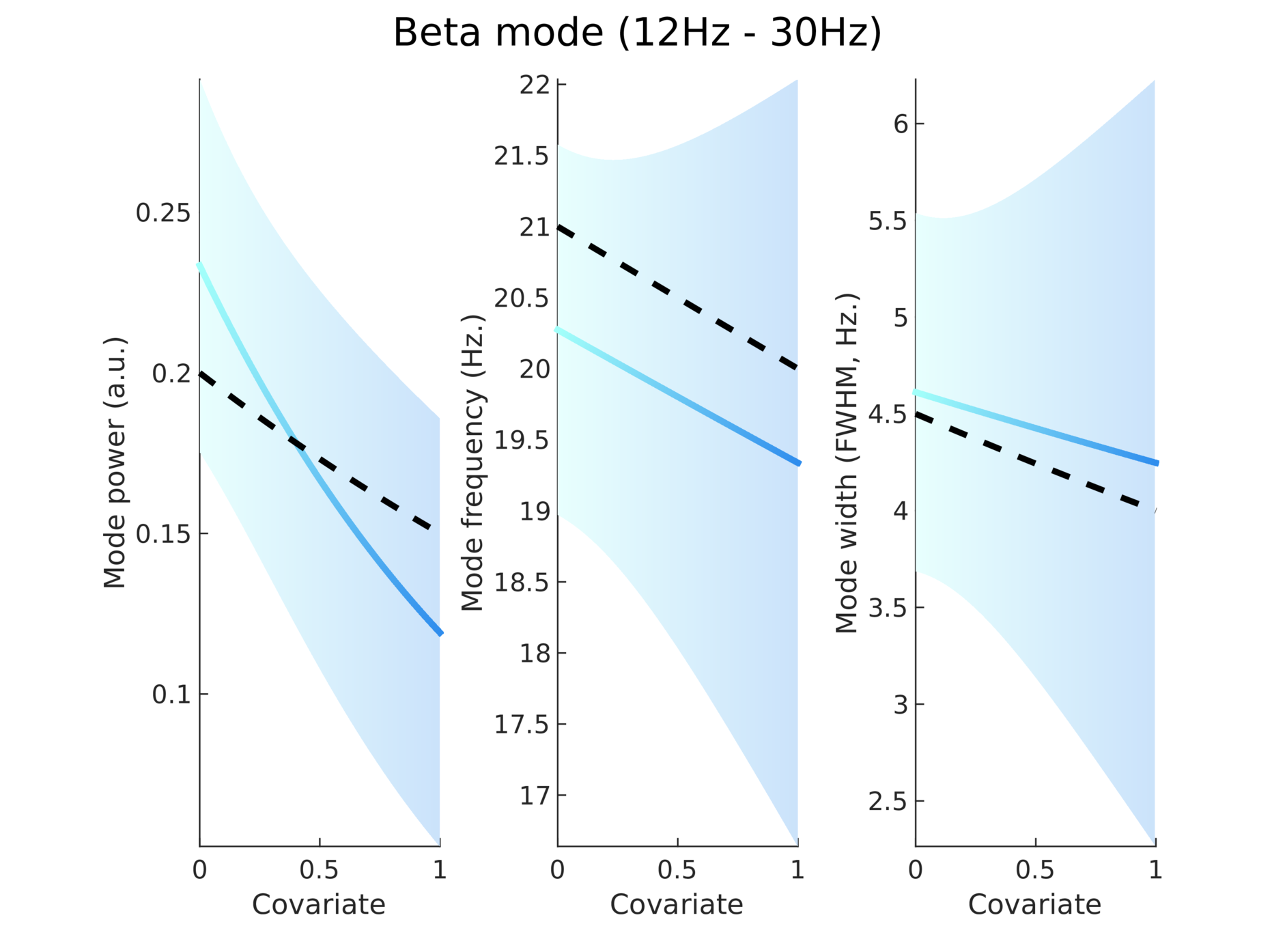}
        \caption{Beta mode during rest}
        \label{fig:sim-peb-beta}
    \end{subfigure}%
    \end{subfigure}
    \caption{a) Design matrix used for second-level analysis, with an intercept as first column and the continuous regressor as second column. b) Recovered spectral density, showing both an effect of the continuous regressor and an effect of the task condition. c) Effect of the continuous regressor on the power (left) and exponent (right) of the aperiodic component. The dashed black line indicates the value that was used to generate the data. The blue line indicates regressed mean of the effect. The shaded area surrounding the mean indicates the 95\% confidence interval. d) Effect of the continuous regressor on the power (left), frequency (middle), and width (right) of the alpha mode. e) Effect of the continuous regressor on the power (left), frequency (middle), and width (right) of the beta mode. }
    \label{fig:sim-peb}
\end{figure*}

%% file: figures/simulation-snrs.tex
\begin{figure*}[t!]
    \centering
    \begin{subfigure}[c]{0.34\textwidth}
        \centering
        \includegraphics[width=\textwidth]{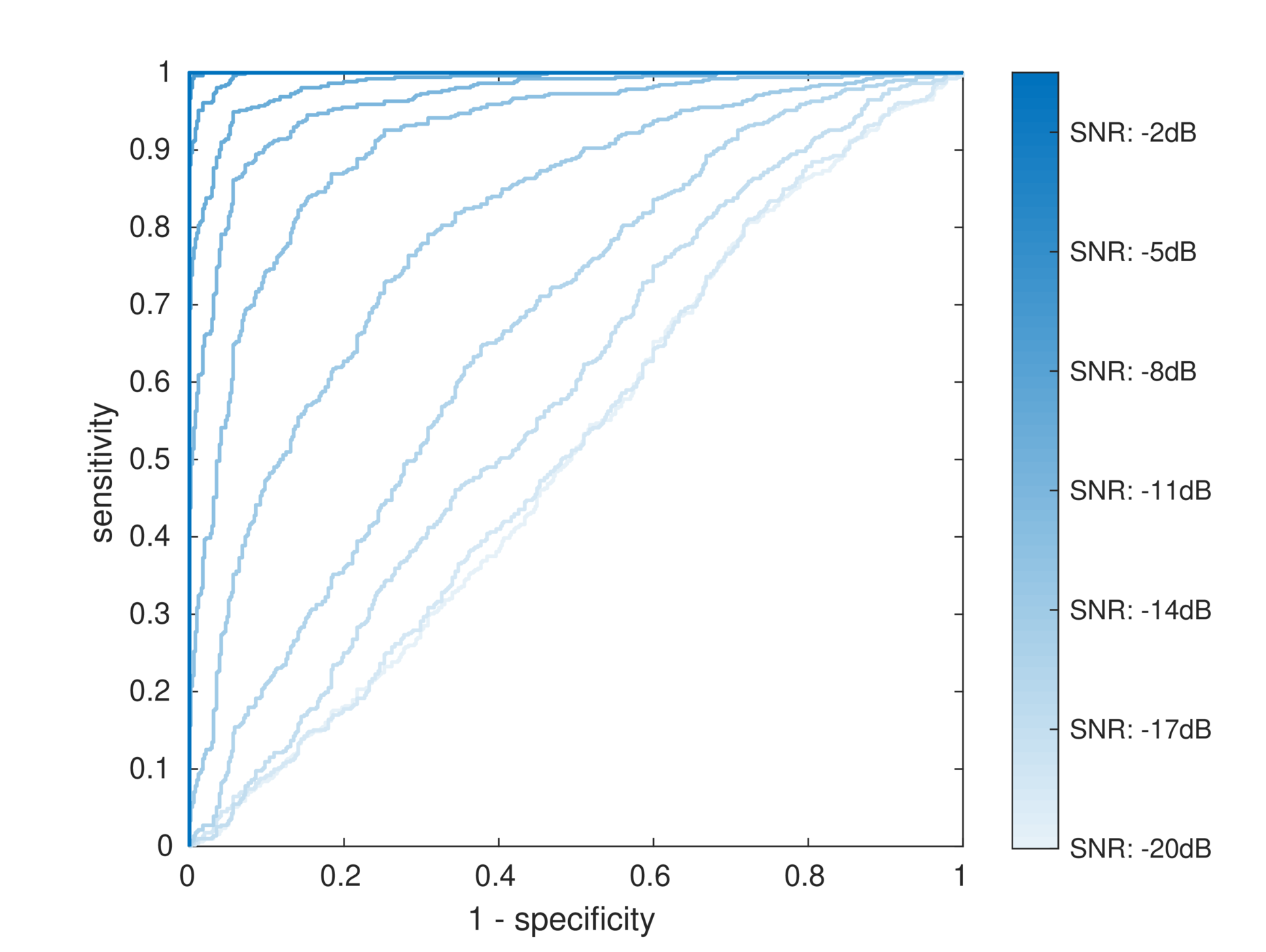}
        \caption{ROC curves at different SNRs}
    \end{subfigure}\hfill
    \begin{subfigure}[c]{0.65\textwidth}
        \centering
        \includegraphics[width=\textwidth]{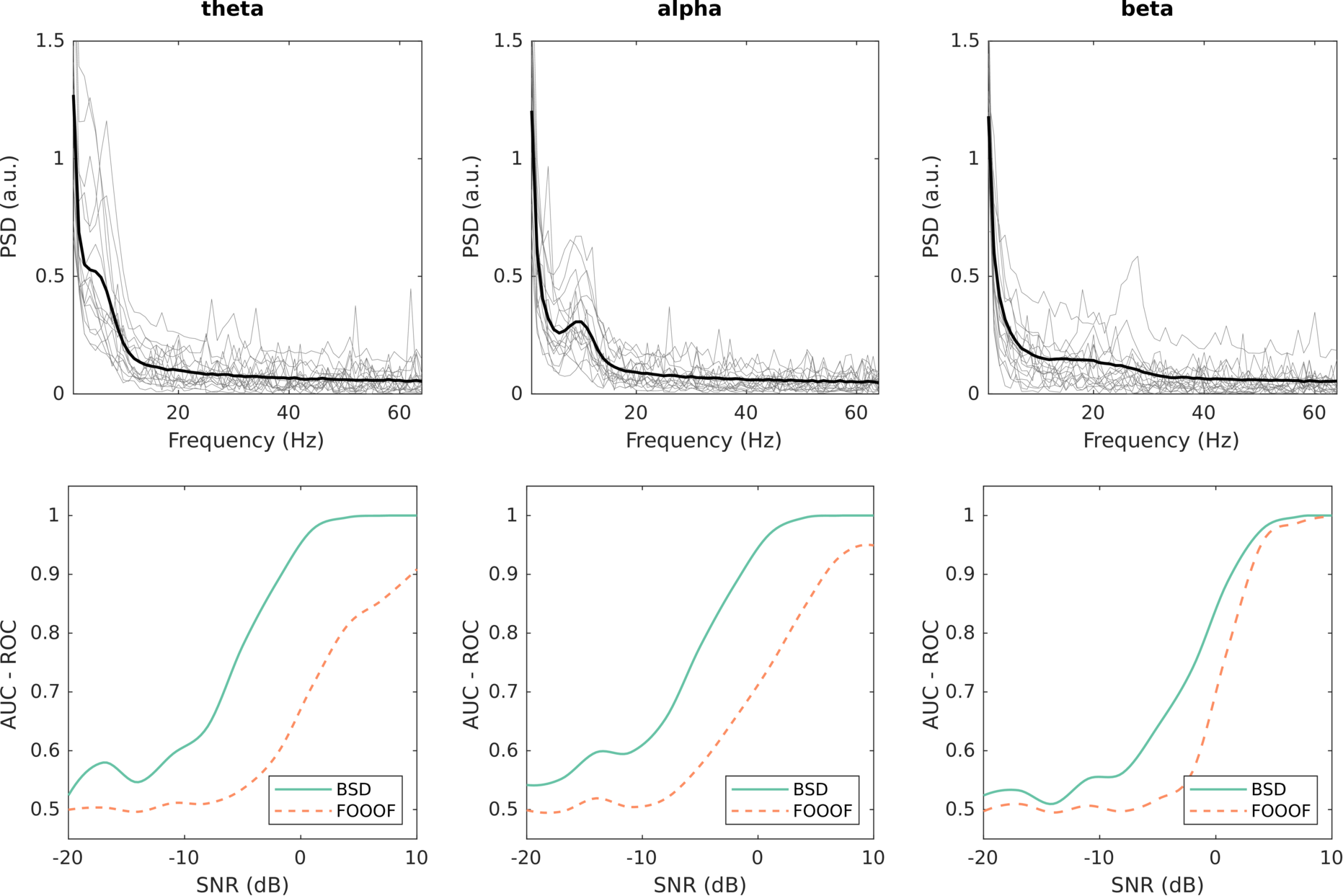}
        \caption{Sample spectrum and detection curves}
    \end{subfigure}
    \caption{(a) ROC curves for detecting a beta peak under  different levels of SNR. (b)  Sample spectrum and detection curves for the theta (left), alpha (middle), and beta (right) frequency bands. The first row shows sample power spectral density for a log precision of the noise of 4 and a SNR of 1~dB. The overlaid black curve shows the sample spectral average. The second row shows the AUC-ROC, i.e. the detection score, as a function of the peak signal-to-noise ratio in decibels for both BSD (in plain green) and FOOOF (in dashed orange). For FOOOF, the default parameters of the official Python implementation (\texttt{specparam}) have been used. }
    \label{fig:sim-snrs}
\end{figure*}

%% file: figures/lemon-bmc.tex
\begin{figure*}[t!]
\centering
\begin{subfigure}[b]{0.6\textwidth}
    \centering
    \includegraphics[width=\textwidth]{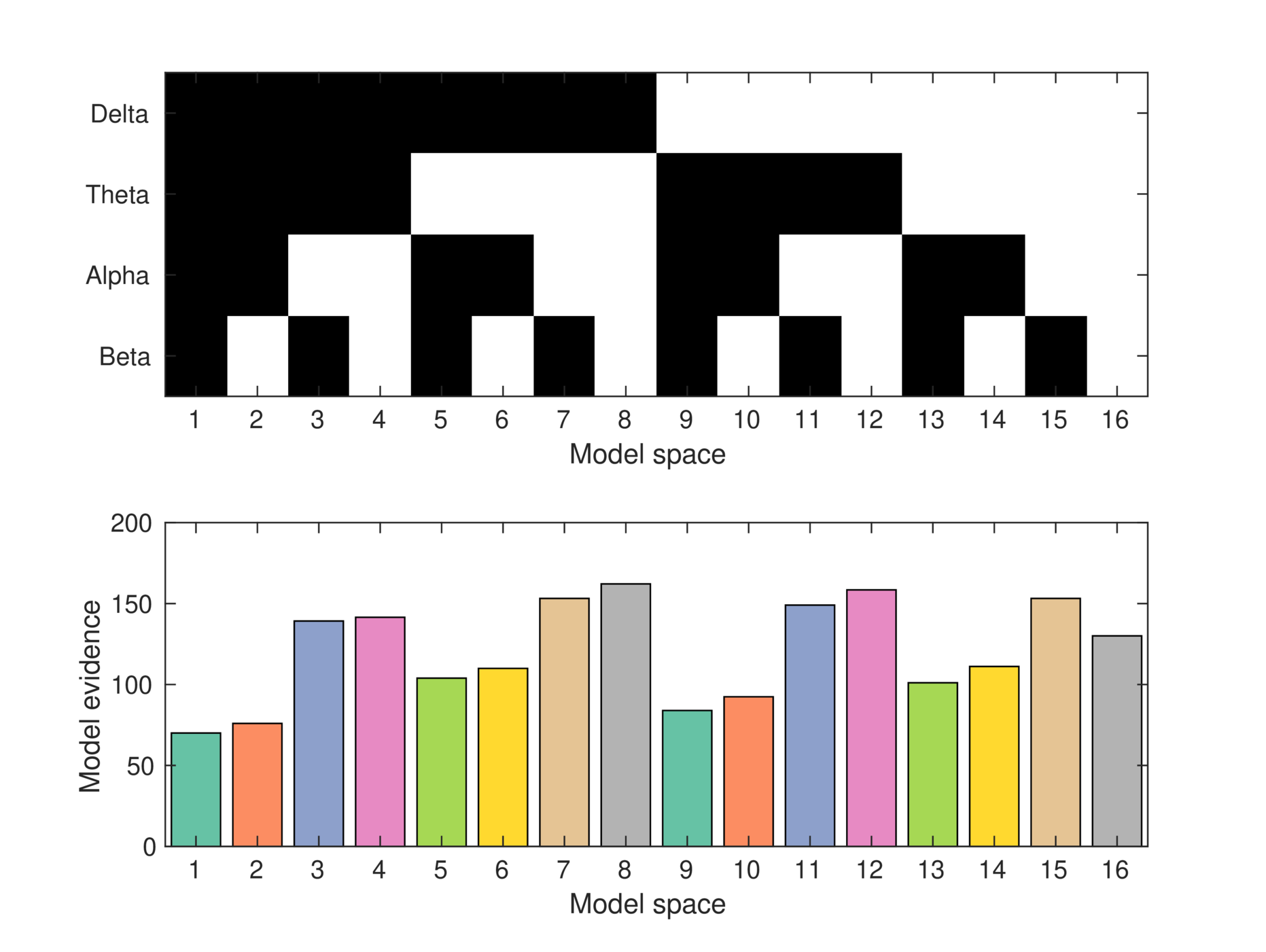}
    \caption{Model space and model evidence}
    \label{fig:lemon-bmc-evidence}
\end{subfigure}
\begin{subfigure}[b]{0.33\textwidth}
    \centering
    \includegraphics[width=\textwidth]{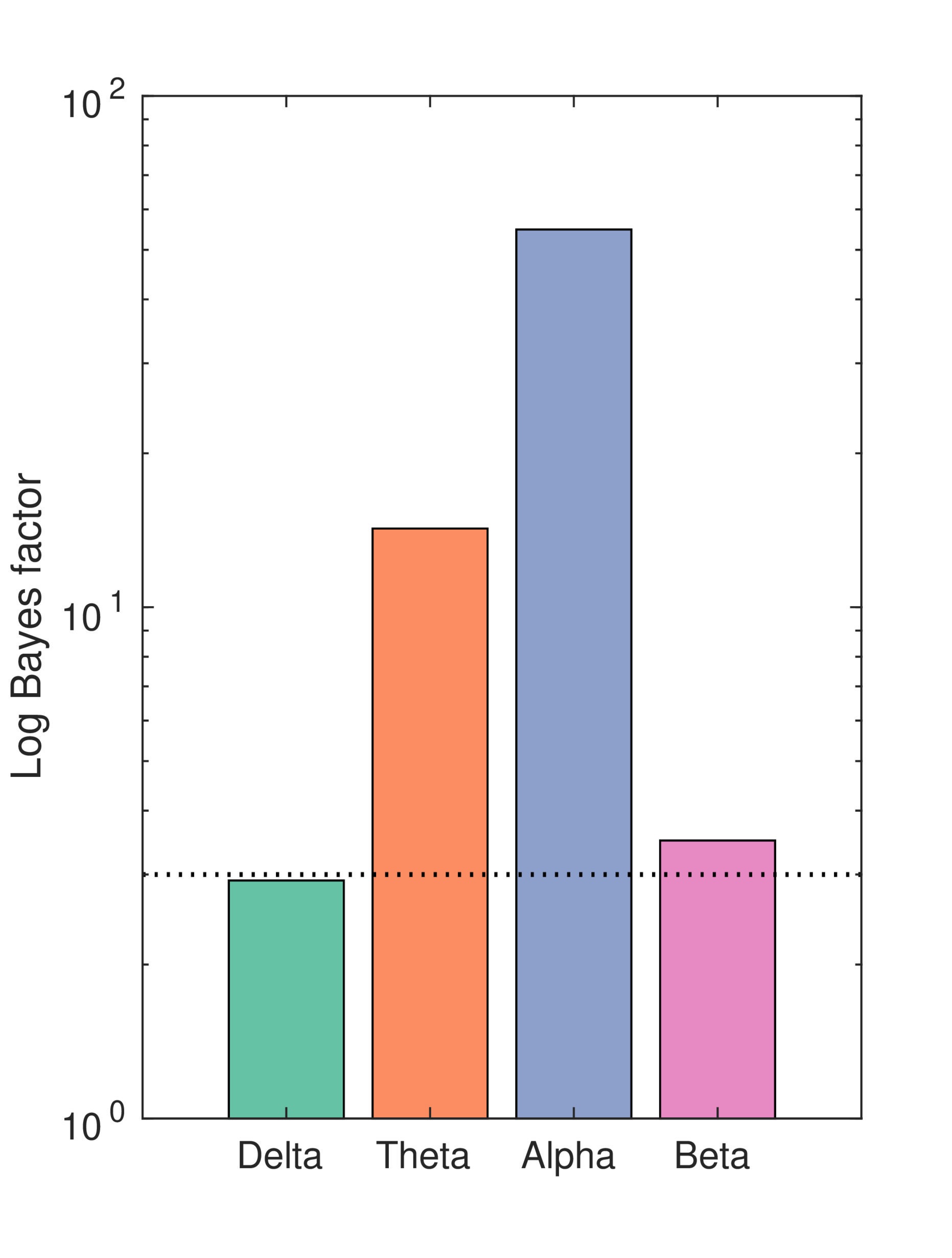} 
    \caption{Log Bayes factor for each peak}
    \label{fig:lemon-bmc-bf}
\end{subfigure}
\caption{Bayesian model comparison results for the LEMON dataset. a) The upper most plot indicates the absence (black) or presence (white) of a particular peak in each model. For instance, a delta peak was present in models 9 to 16. The lower most plot indicates the model evidence, as estimated by the free-energy, for each of the 16 models. b) Log Bayes factors for the presence of each peak. These are obtained by averaging the free-energy over models that feature a particular peak, and subtracting the average free-energy over models without that peak. The dotted horizontal lines indicate a Bayes factor of 3, commonly accepted a threshold for strong evidence. Both theta and alpha peak have decisive evidence, while beta peak only has strong evidence. }
\label{fig:lemon-bmc}
\end{figure*}

%% file: figures/lemon-overall.tex
\begin{figure*}[t!]
\centering
\begin{subfigure}[b]{0.49\textwidth}
    \centering
    \includegraphics[width=\textwidth]{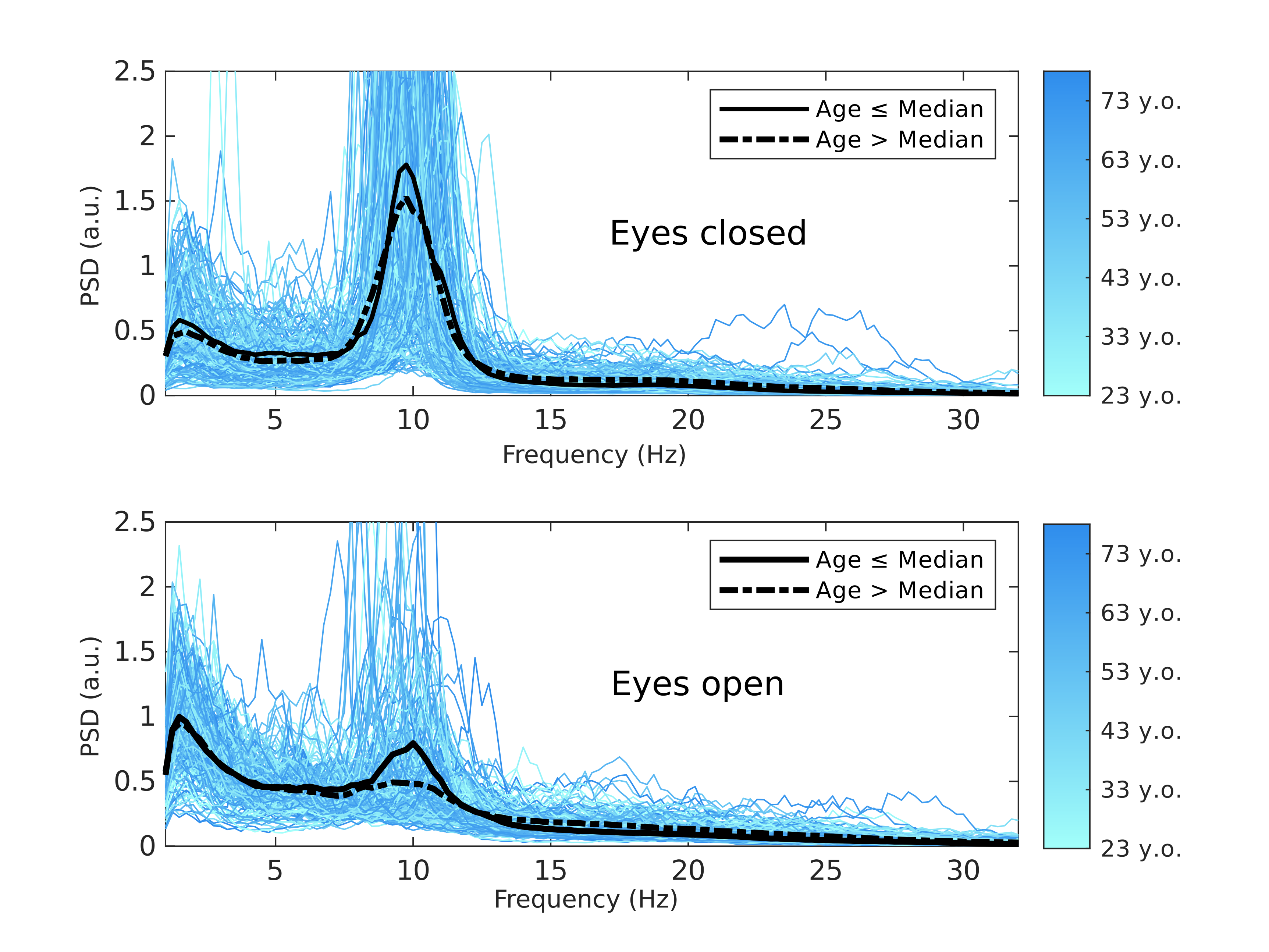}
    \caption{Observed data}
    \label{fig:lemon-observed}
\end{subfigure}%
\begin{subfigure}[b]{0.49\textwidth}
    \centering
    \includegraphics[width=\textwidth]{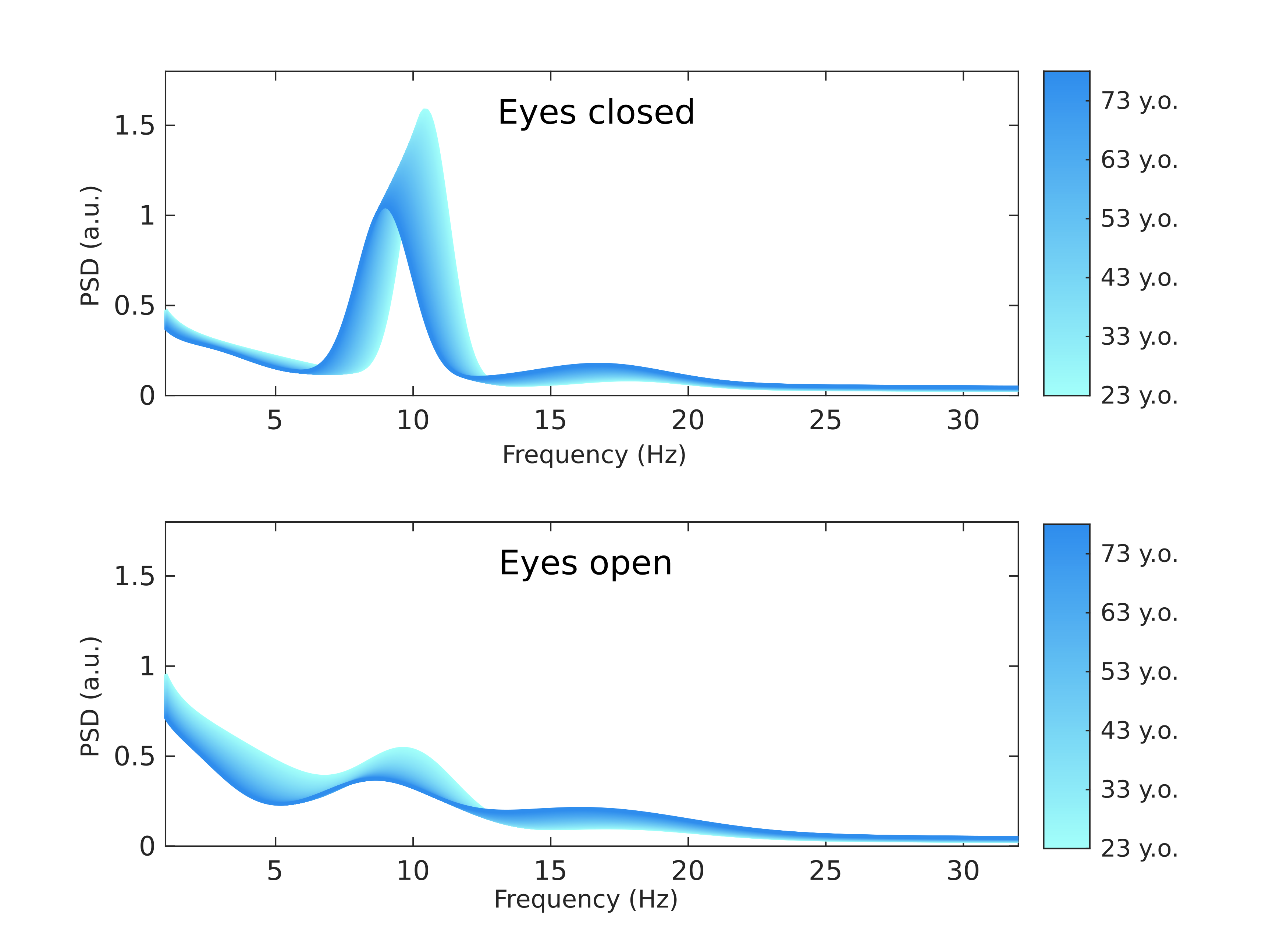}
    \caption{Recovered model}
    \label{fig:lemon-recovered}
\end{subfigure}
\caption{Observed spectrum and estimated model for the LEMON dataset. a) Sample spectra in both eyes closed and eyes opened conditions for the 202 subjects. The plain black line shows the spectral average for subjects with a continuous regressor value below the population median, and the dashed line corresponds to the average for subjects above the median. The color gradient indicates the age, with light cyan indicating 22.5 years old to dark blue indicating 77.5 years old. b) Ribbon plot of the generative model, i.e., the model recovered from the observed data in both rest and task conditions. The color gradient indicates the age. }
\label{fig:lemon-overall}
\end{figure*}

%% file: figures/lemon-modes.tex
\begin{figure*}[t!]
\centering
\begin{subfigure}{\textwidth}
\begin{subfigure}[b]{0.49\textwidth}
    \centering
    \includegraphics[width=0.9\textwidth]{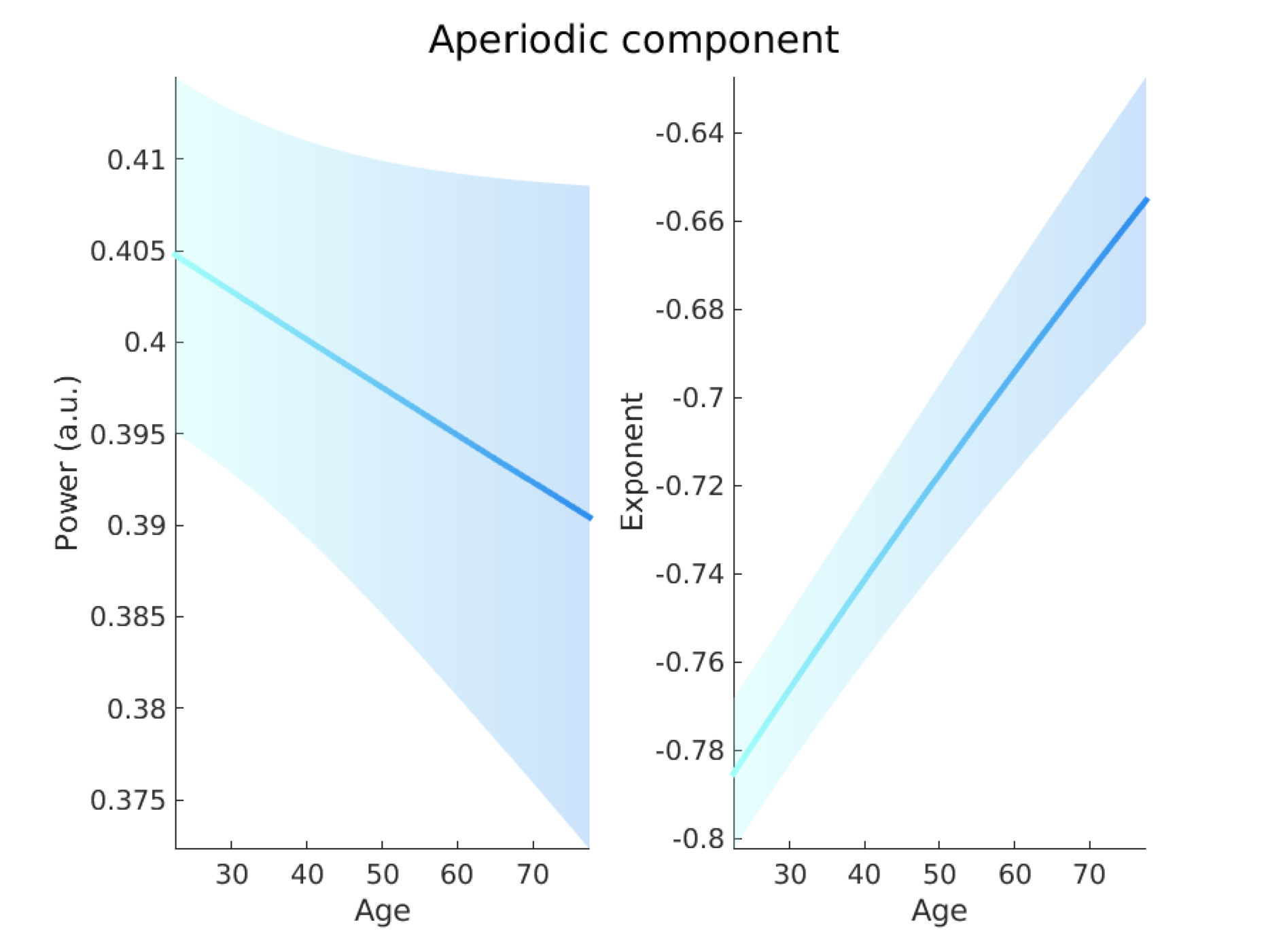}
    \caption{Effect of age on the aperiodic component }
    \label{fig:lemon-aperiodic}
\end{subfigure}
\begin{subfigure}[b]{0.49\textwidth}
    \centering
    \includegraphics[width=0.9\textwidth]{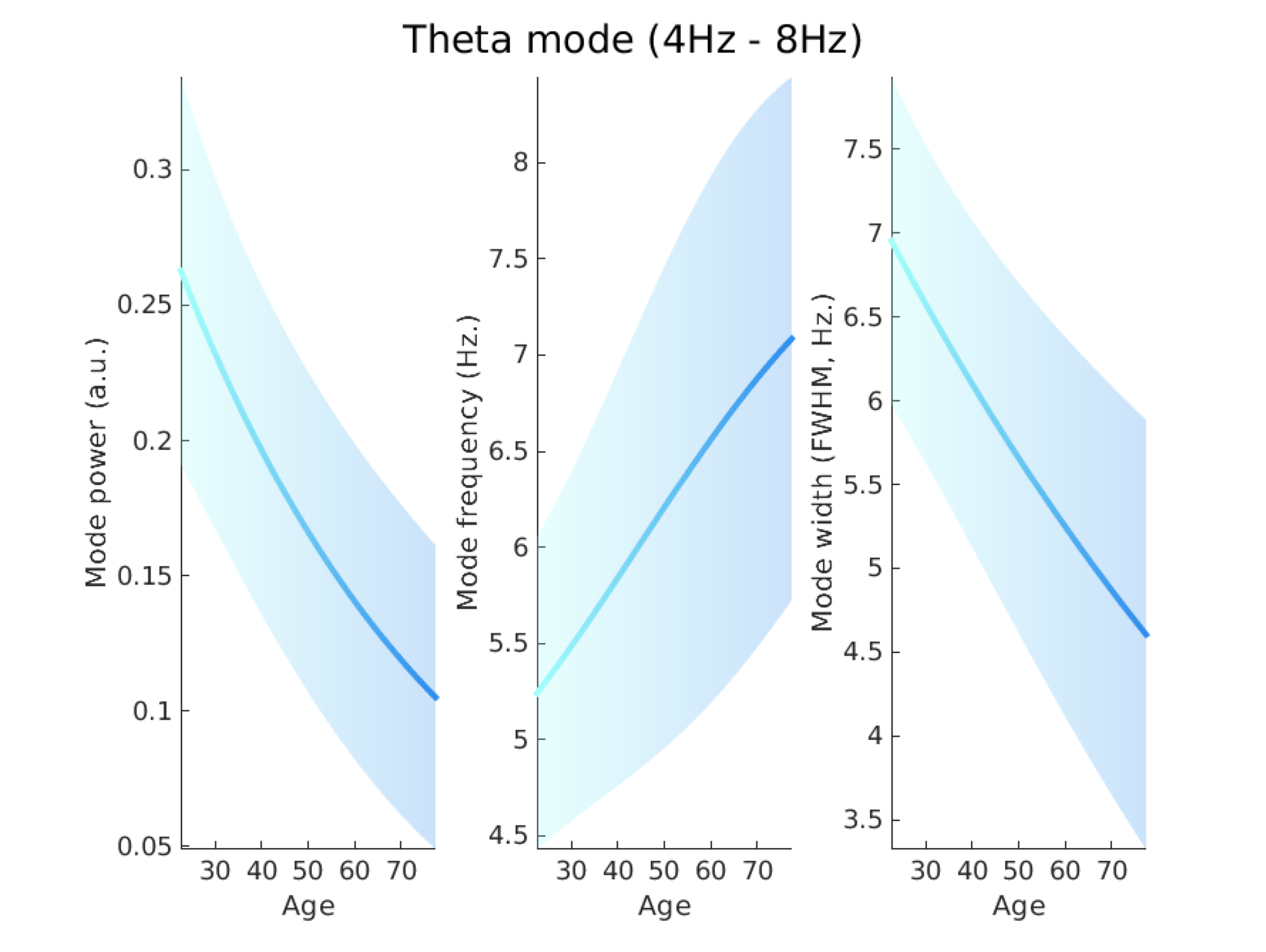} 
    \caption{Effect of age on the theta mode}
    \label{fig:lemon-theta}
\end{subfigure}
\vspace{3em}
\end{subfigure}
\begin{subfigure}{\textwidth}
\begin{subfigure}[b]{0.49\textwidth}
    \centering
        \includegraphics[width=0.9\textwidth]{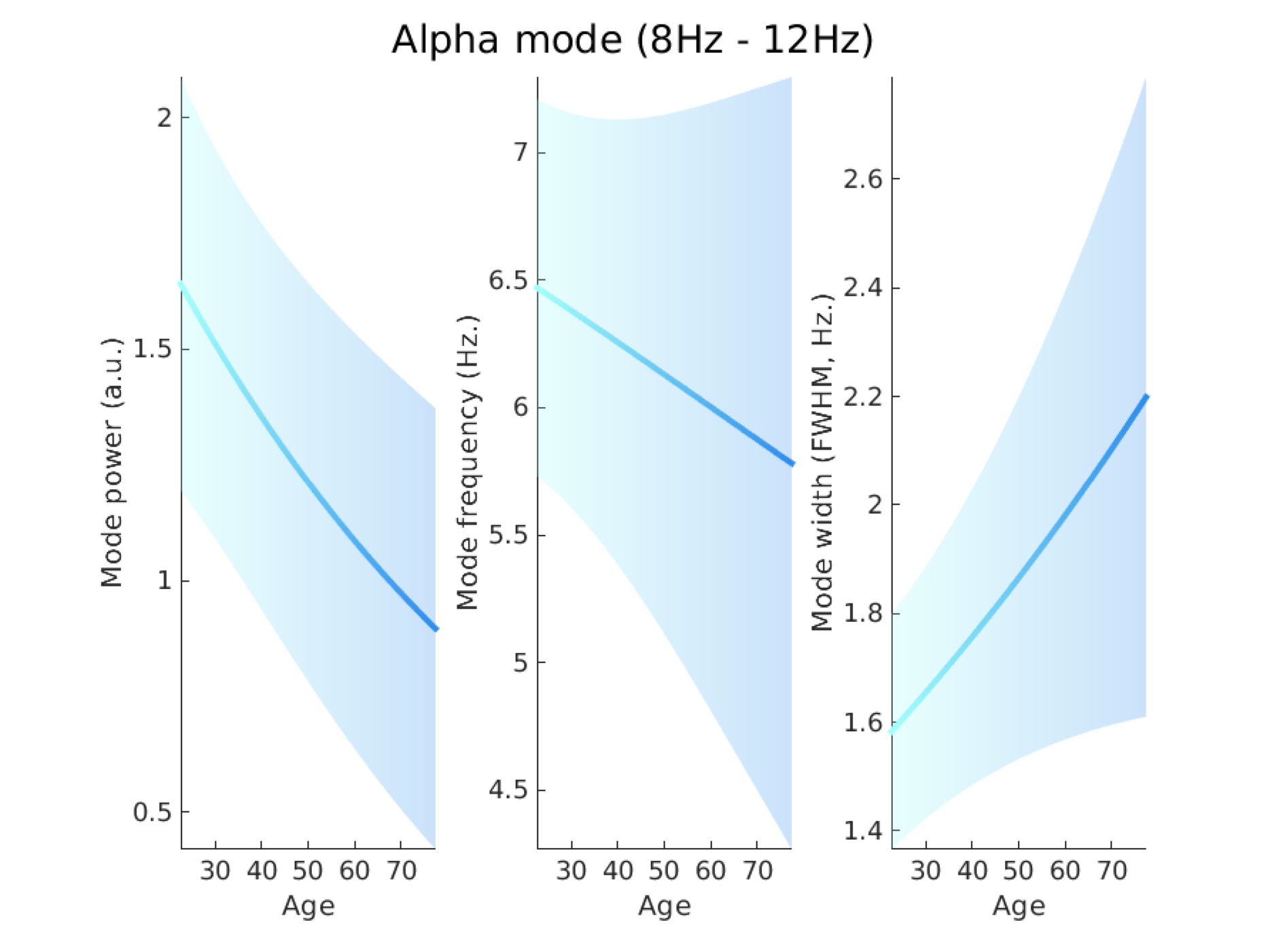}
    \caption{Effect of age on the alpha mode}
    \label{fig:lemon-alpha}
\end{subfigure}%
\begin{subfigure}[b]{0.49\textwidth}
    \centering
    \includegraphics[width=0.9\textwidth]{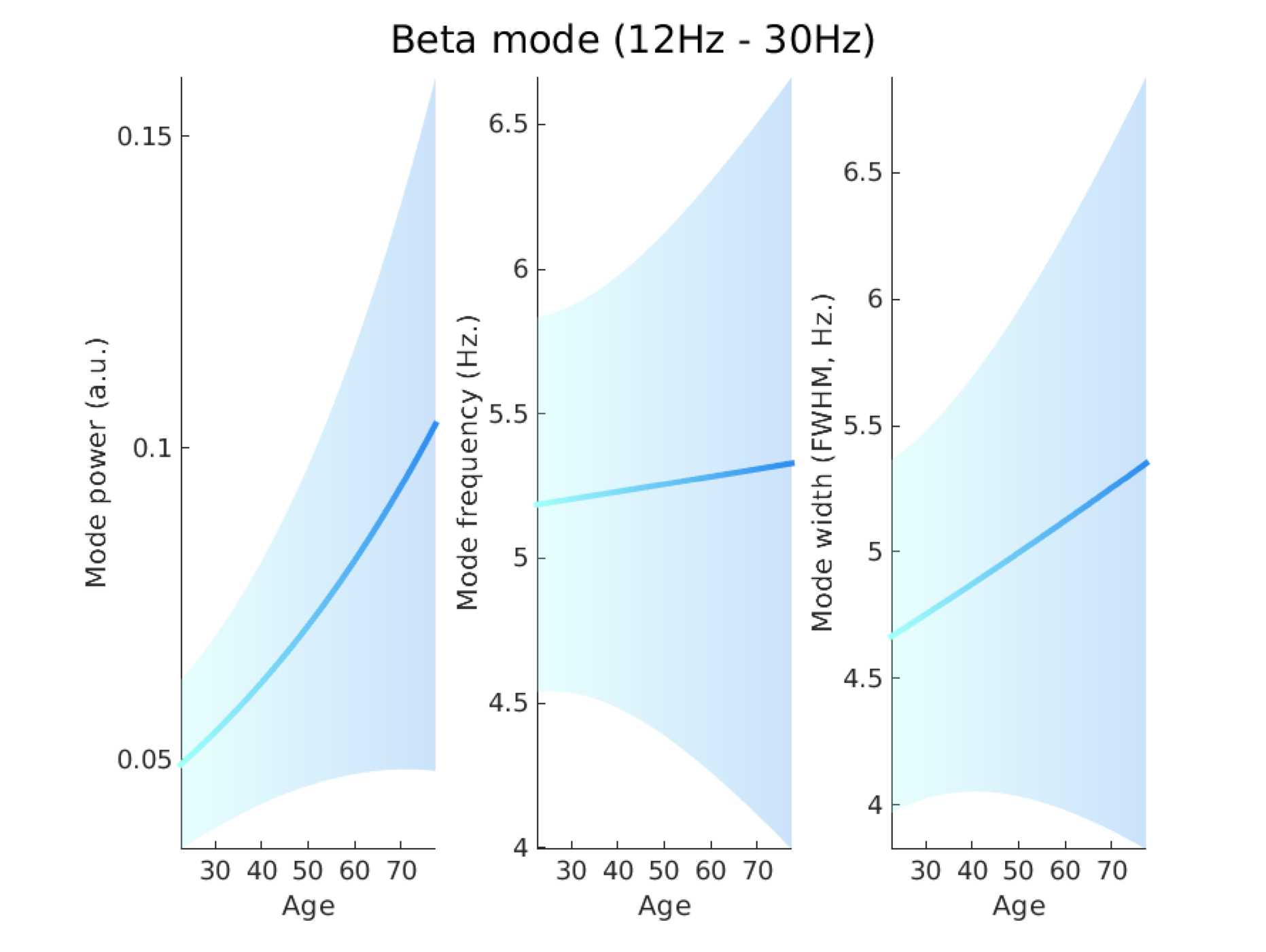}
    \caption{Effect of age on the beta mode}
    \label{fig:lemon-beta}
\end{subfigure}
\end{subfigure}
\caption{Effect of age on model parameters of the aperiodic component (a) and theta (b), alpha (c), and beta modes (d). For each plot, the blue line indicates regressed mean of the effect, whilst the shaded area surrounding the mean indicates the 90\% confidence interval.  }
\label{fig:lemon-modes}
\end{figure*}